\documentclass[%
 reprint,
 amsmath,amssymb,
prx,
floatfix,
superscriptaddress
]{revtex4-2}

\usepackage{mathtools}
\usepackage{graphicx}
\usepackage{dcolumn}
\usepackage{bm}

\usepackage[caption=false]{subfig}

\usepackage{algorithm}
\usepackage[noend]{algpseudocode}
\algrenewcommand\algorithmicdo{}

\makeatletter
\renewcommand{\ALG@name}{Procedure}
\makeatother

\newcommand{\switch}{%
  \mathcode`+=\numexpr\mathcode`+ + "1000\relax 
  \mathcode`*=\numexpr\mathcode`* + "1000\relax
}

\usepackage[linktocpage=true,
  colorlinks=true, 
  pdfborder={0 0 0},
  linkcolor=blue,
  citecolor=red,
  filecolor=yellow,
  urlcolor=blue,
  bookmarks,
  pdfauthor={},
]{hyperref}

\usepackage{orcidlink}

\usepackage{ifthen}
\newcounter{is_qcircuit_used}
\newcounter{are_figs_merged}
\newcommand{\argmax}{\mathop{\rm arg~max}\limits}

\begin{document}

\preprint{APS/123-QED}

\title{
Qubit encoding for a mixture of localized functions
}

\author{Taichi Kosugi\orcidlink{0000-0003-3379-3361}}
\email{kosugi.taichi@gmail.com}
\affiliation{
Laboratory for Materials and Structures,
Institute of Innovative Research,
Tokyo Institute of Technology,
Yokohama 226-8503,
Japan
}

\affiliation{
Quemix Inc.,
Taiyo Life Nihombashi Building,
2-11-2,
Nihombashi Chuo-ku, 
Tokyo 103-0027,
Japan
}

\affiliation{
Department of Physics,
The University of Tokyo,
Tokyo 113-0033,
Japan
}

\author{Shunsuke Daimon\orcidlink{0000-0001-6942-4571}}
\affiliation{Quantum Materials and Applications Research Center,
National Institutes for Quantum Science and Technology (QST),
2-12-1 Ookayama, Meguro-ku, Tokyo 152-8550, Japan
}

\author{Hirofumi Nishi\orcidlink{0000-0001-5155-6605}}
\affiliation{
Laboratory for Materials and Structures,
Institute of Innovative Research,
Tokyo Institute of Technology,
Yokohama 226-8503,
Japan
}

\affiliation{
Quemix Inc.,
Taiyo Life Nihombashi Building,
2-11-2,
Nihombashi Chuo-ku, 
Tokyo 103-0027,
Japan
}

\affiliation{
Department of Physics,
The University of Tokyo,
Tokyo 113-0033,
Japan
}

\author{Shinji Tsuneyuki\orcidlink{0009-0004-8790-7429}}
\affiliation{
Department of Physics,
The University of Tokyo,
Tokyo 113-0033,
Japan
}

\author{Yu-ichiro Matsushita\orcidlink{0000-0002-9254-5918}}
\affiliation{
Laboratory for Materials and Structures,
Institute of Innovative Research,
Tokyo Institute of Technology,
Yokohama 226-8503,
Japan
}

\affiliation{
Quemix Inc.,
Taiyo Life Nihombashi Building,
2-11-2,
Nihombashi Chuo-ku, 
Tokyo 103-0027,
Japan
}

\affiliation{Quantum Materials and Applications Research Center,
National Institutes for Quantum Science and Technology (QST),
2-12-1 Ookayama, Meguro-ku, Tokyo 152-8550, Japan
}

\affiliation{
Department of Physics,
The University of Tokyo,
Tokyo 113-0033,
Japan
}

\date{\today}

\begin{abstract}
One of the crucial generic techniques for quantum computation is amplitude encoding.
Although several approaches have been proposed,
each of them often requires exponential classical-computational cost
or an oracle whose explicit construction is not provided.
Given the growing demands for practical quantum computation,
we develop moderately specialized encoding techniques
that generate an arbitrary linear combination of localized complex functions.
We demonstrate that $n_{\mathrm{loc}}$ discrete Lorentzian functions as an expansion basis set lead to efficient probabilistic encoding,
whose computational time is 
$
\mathcal{O}
( \max (
n_{\mathrm{loc}}^2 \log n_{\mathrm{loc}},
n_{\mathrm{loc}}^2 \log n_q, n_q
))
$
for $n_q$ data qubits equipped with $\log_2 n_{\mathrm{loc}}$ ancillae.
Furthermore, amplitude amplification in combination with amplitude reduction renders it deterministic analytically
with controllable errors and the computational time is reduced to
$
\mathcal{O}
( \max (
n_{\mathrm{loc}}^{3/2} \log n_{\mathrm{loc}}, 
n_{\mathrm{loc}}^{3/2} \log n_q, n_q 
)).
$
We estimate required resources for applying our scheme to quantum chemistry in real space.
We also show the results on real superconducting quantum computers to confirm the validity of our techniques.
\end{abstract}

\maketitle 

\section{Introduction}
\label{sec:introduction}

We often encounter a case where a quantum algorithm that is mathematically proven to solve a problem more efficiently than classical algorithms assumes that an initial many-qubit state has already been prepared in which the initial condition is appropriately encoded in some format specified by the algorithm.
Such generic techniques for preparing an arbitrary many-qubit state from initialized qubits as efficiently as possible are called amplitude encoding.
Since an $n$-qubit system has approximately $2^n$ degrees of freedom,
encoding a truly arbitrary state, i.e., a state under no constraints, using predetermined circuit parameters inevitably suffers from exponential classical-computational cost with respect to $n$ \cite{bib:5694, bib:5695, bib:5693, bib:5336, bib:5692, bib:5698}.
If we exploit quantum parallelism, on the other hand,
exponential cost can be avoided.
There are efficient encoding techniques for arbitrary states under no or only weak constraints \cite{bib:5394, bib:5705, bib:5605}.
Each of them assumes, however, that an oracle capable of performing an elaborate unitary operation on $\mathcal{O} (2^n)$ independent states simultaneously is available.
This fact prohibits us from finding an efficient encoding scheme for practical use today.

When we tackle a specific problem, however, such techniques for arbitrary states are often found to be too generic; that is,
plausible initial states for the problems are characterized by much fewer parameters than $2^n.$
Let us consider, for example, a quantum algorithm that admits an initial state in which a probability distribution function is encoded.
Even if we want to try various normal distributions as the inputs,
a generic encoding technique demands exponential cost,
despite the states being specified only by their means and widths.
Intuitively,
the degrees of freedom in a generic encoding technique are unnecessarily enormous compared to the amount of information specifying an initial state of practical use.
Motivated by this consideration,
we develop moderately specialized encoding techniques in the present study: generating an arbitrary linear combination (LC) of localized complex functions as a fair trade-off between the degrees of freedom and the computational cost.

To this end, we design the main encoding technique by starting from the probabilistic operation \cite{bib:5005, bib:5163} for an LC of discrete Lorentzian functions (LFs),
to which we apply the quantum amplitude amplification (QAA) technique \cite{bib:4884, bib:4878} to render the encoding deterministic.
As is demonstrated later,
expanding a target function in LFs is favorable for achieving efficient circuit implementation.
Our encoding techniques possess applicability to diverse fields of quantum computation.
Amongst them, quantum chemistry in real space \cite{bib:5373, bib:5372, bib:5328, bib:5824, bib:5737, bib:5658, bib:6103, bib:6236, bib:6242, bib:6455, bib:Horiba} is a promising one,
where an initial state can be constructed from one-electron molecular orbitals (MOs) \cite{bib:4825, bib:5389},
typically expressed as LCs of localized orbitals.
We provide resource estimation for such calculations.
To demonstrate the efficiency and practical usefulness of our scheme,
we perform it on the cloud quantum computing platform commercially provided by IBM \cite{ibmq_quantum}.
Our encoding scheme is for a dense quantum state;
that is, it treats a target function that has basically a nonzero amplitude of each computational basis.
For encoding schemes for a sparse quantum state,
see, e.g., Refs \cite{bib:6239, bib:5767, bib:5646} and references therein.

It is noted that encoding techniques exist based on training of parametrized circuits \cite{bib:5706, bib:5688}.
Although such techniques are promising for practical use,
we focus on those based on something other than training in the present study.
Also, we note here that Moosa {\it et al.}\cite{bib:6136} proposed the Fourier series loader method.
It encodes first the Fourier coefficients of a target function on a coarse mesh,
after which they are reinterpreted as those for a finer mesh to be Fourier transformed backward.
See also Ref.~\cite{bib:5929} for the quantum interpolation method.

\section{Methods}

\subsection{Determinization of the probabilistic operation for an LC of unitaries}
\label{sec:generic_determinization}

\subsubsection{Probabilistic operation of an LC of unitaries and amplitude reduction}

\begin{figure*}
\begin{center}
\includegraphics[width=15cm]{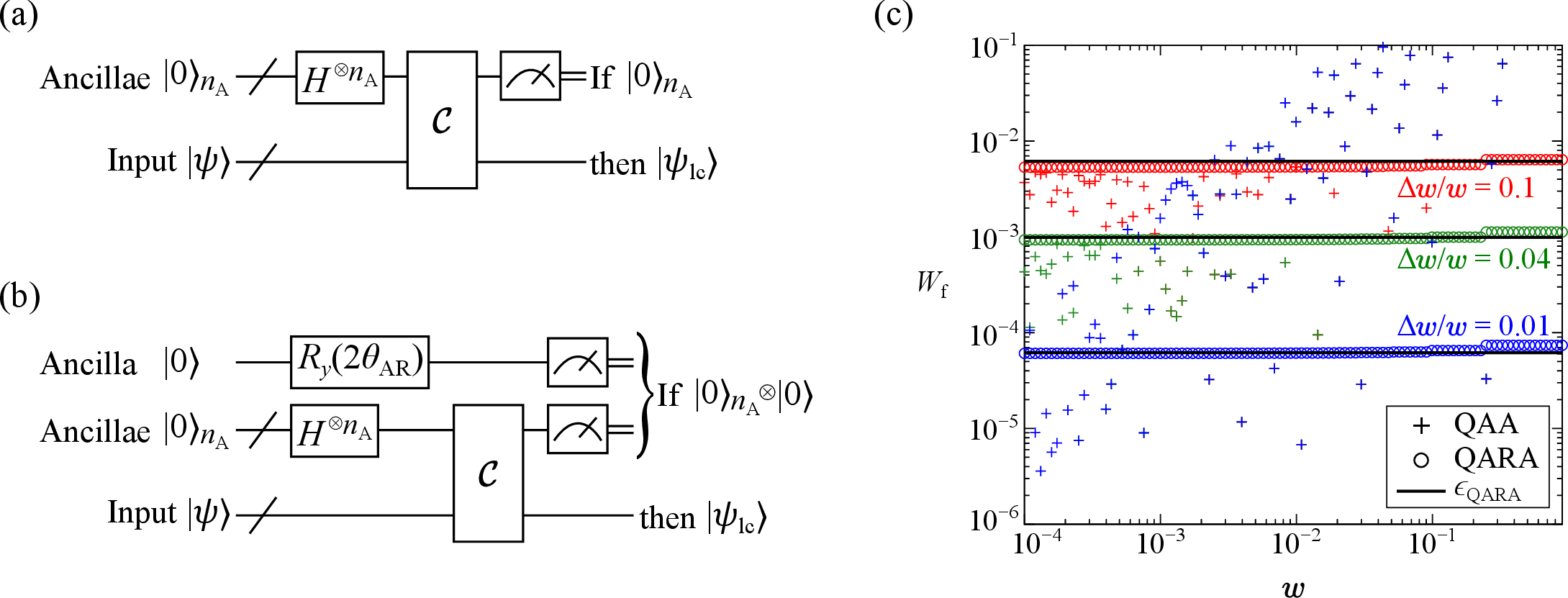}
\end{center}
\caption{
Each of circuits (a) $\mathcal{C}_{\mathrm{prob}}$ and (b) $\mathcal{C}_{\mathrm{prob}}^{(\mathrm{AR})}$ is for probabilistic operation of an LC $\mathcal{A}$ of unitaries.
The explicit construction of the central part $\mathcal{C}$ in $\mathcal{C}_{\mathrm{prob}}$ is described in the original paper \cite{bib:5163}.
$\mathcal{C}_{\mathrm{prob}}^{(\mathrm{AR})}$ implements the amplitude reduction for the success state by introducing an extra ancilla in addition to those in $\mathcal{C}_{\mathrm{prob}}.$
(c)
Each circle represents the weight $W_{\mathrm{f}}$ of failure state in the resultant state of QARA with a QAE error $\Delta w$ as a function of the true weight $w$ of success state in the original state.
Each cross represents $W_{\mathrm{f}}$ for QAA-only with a QAE error.
Horizontal lines represent the approximate maximum errors in
Eq.~(\ref{error_max_of_QARA}) for the values of $\Delta w/w.$ 
}
\label{fig:lin_combo_of_unitaries_and_qara}
\end{figure*}

For an arbitrary many-qubit state $| \psi \rangle,$
it has already been demonstrated to be possible to
operate an arbitrary LC $\mathcal{A}$ of $n_{\mathrm{loc}}$ unitaries on $| \psi \rangle$ by using the circuit $\mathcal{C}_{\mathrm{prob}},$ as depicted in
Fig.~\ref{fig:lin_combo_of_unitaries_and_qara}(a) \cite{bib:5005, bib:5163}.
$\mathcal{A}$ is nonunitary in general.
$\mathcal{C}_{\mathrm{prob}}$ uses
$n_{\mathrm{A}} = \log_2 n_{\mathrm{loc}}$
(assumed to be an integer for simplicity)
ancillae to implement $\mathcal{A}$ probabilistically.
The central part $\mathcal{C}$ of the circuit consists of the multiply controlled constituent unitaries and the rotation gates on the ancillae.
For details, see the original paper \cite{bib:5163}.
The state of the whole system immediately before the measurement is written in the form
\begin{align}
    | \Psi \rangle
    =
        \sqrt{w}
        | \psi_{\mathrm{lc}} \rangle
        \otimes
        | 0 \rangle_{n_{\mathrm{A}} }
        +
        \sum_{j = 1}^{ 2^{n_{\mathrm{A}}} - 1}
            \sqrt{w_j}
            | \psi_j \rangle
            \otimes
            | j \rangle_{n_{\mathrm{A}} }
    ,
    \label{ampl_reduction:state_before_meas}
\end{align}
where $w$ is the weight of the desired state
$| \psi_{\mathrm{lc}} \rangle \propto \mathcal{A} | \psi \rangle$
and $w_j$ is that of the undesired state coupled to the $j$th computational basis of the ancillae.
The desired state is obtained only when the observed state is
$| 0 \rangle_{n_{\mathrm{A}}}.$
The success probability is $w,$
equal to the weight of the desired state in $| \Psi \rangle$ by definition.
Since this scheme begins with assigning the equal weights to the computational bases for the ancillae,
the success probability is on the order of
$1/2^{n_{\mathrm{A}}} = 1/n_{\mathrm{loc}}.$

When the original technique of QAA \cite{bib:4884, bib:4878} is applied to $\mathcal{C}_{\mathrm{prob}},$
the success probability is made close to 1.
The amplified value is, however, not necessarily equal to 1, depending on the original success probability.
This fact motivates us to define another circuit 
$\mathcal{C}_{\mathrm{prob}}^{(\mathrm{AR})}$
for the probabilistic operation of $\mathcal{A}.$
This circuit differs from 
$\mathcal{C}_{\mathrm{prob}}$ only in that
the new one uses an extra ancilla, as depicted in 
Fig.~\ref{fig:lin_combo_of_unitaries_and_qara}(b).
The extra ancilla is accompanied by the $y$ rotation gate
$R (2 \theta_{\mathrm{AR}}) = \exp (- i \theta_{\mathrm{AR}} \sigma_y)$
with an angle parameter $\theta_{\mathrm{AR}}.$
Although $\mathcal{C}_{\mathrm{prob}}^{(\mathrm{AR})}$
involves $\mathcal{C}_{\mathrm{prob}}$ as its part separated from the extra ancilla,
they will be coupled for QAA later. 
The state $| \Psi_{\mathrm{AR}} \rangle$ of the whole system immediately before the measurement is obviously the tensor product of $| \Psi \rangle$
in Eq.~(\ref{ampl_reduction:state_before_meas})
and the extra-ancilla state
$
\cos \theta_{\mathrm{AR}}
| 0 \rangle
+
\sin \theta_{\mathrm{AR}}
| 1 \rangle
.
$
We regard only
$| 0 \rangle_{n_{\mathrm{A}} } \otimes | 0 \rangle$
to be the success state,
even though
$| 0 \rangle_{n_{\mathrm{A}} } \otimes | 1 \rangle$
also leads to $| \psi_{\mathrm{lc}} \rangle.$
The success probability is thus given by
\begin{align}
    w_{\mathrm{AR}}
    =
        w
        \cos^2
        \theta_{\mathrm{AR}}
    ,
    \label{gen_loc_state:success_prob_after_ampl_reduction}
\end{align}
which means that we can reduce the success probability $w$ for the old circuit by an arbitrary factor with an appropriate $\theta_{\mathrm{AR}}$ for the new circuit.
We refer to this treatment as the amplitude reduction for probabilistic operation.

\subsubsection{Determinization via amplitude reduction and amplification}

The generic technique of QAA \cite{bib:4884, bib:4878} amplifies the weight of `good' states contained in an input state by rotating it within the two-dimensional subspace spanned by the good states and the orthogonal states,
leading to the quadratic speedup for obtaining the good state.
This rotation is achieved by the appropriate number of applications of the amplification operator to the input state.
In the present study, we adopt this technique for
$| \Psi_{\mathrm{AR}} \rangle$
in $\mathcal{C}_{\mathrm{prob}}^{(\mathrm{AR})}$
to raise the success probability.
The good states in this case are all those in which the ancillary part is $| 0 \rangle_{n_{\mathrm{A}} + 1}.$
The amplification operator $Q$ for this case can be implemented according to the generic theory.
We assume that we already know the weight $w$ of the desired state $| \psi_{\mathrm{lc}} \rangle$ in $\mathcal{C}_{\mathrm{prob}}$
precisely by employing techniques of quantum amplitude estimation (QAE)\cite{bib:4878, bib:5145, bib:5594, bib:5655, bib:5996, bib:5997, bib:6185}.
Let us consider $m$ applications of $Q$ to
$| \Psi_{\mathrm{AR}} \rangle$ by appending the amplification circuit to $\mathcal{C}_{\mathrm{prob}}^{(\mathrm{AR})}.$
By defining $\theta_w$ via $w = \sin^2 \theta_w \ (0 < \theta_w < \pi/2),$
the weight of the good state after the amplification is written as
$w_m (\theta_w) \equiv \sin^2 ((2 m + 1) \theta_w)$ \cite{bib:4884, bib:4878}.
If $w_m (\theta_w)$ is equal to unity,
the desired state will be obtained with certainty;
that is, the nonunitary operation of $\mathcal{A}$ is now deterministic.
The operation is still probabilistic otherwise.
Specifically, for a fixed $m,$
the following $m$ values of $\theta_w,$ 
\begin{align}
    \theta^{(m, n)}
    \equiv
        \frac{2 n + 1}{4 m + 2}
        \pi
        \
        (n = 0, \dots, m - 1)
    ,
    \label{gen_loc_state:theta_m_n_for_QAA}
\end{align}
lead to $w_m (\theta^{(m, n)}) = 1.$
If $\theta_w$ is fortunately equal to any one of them,
$Q^m$ makes the weight of the good state unity.
By choosing a sufficiently large $m,$
we can let the smallest value
$\theta^{(m, 0)} = \pi/(4 m + 2)$
among those in Eq.~(\ref{gen_loc_state:theta_m_n_for_QAA})
be smaller than $\theta_w.$
The smallest $m$ that satisfies $\theta^{(m, 0)} < \theta_w$ is
\begin{align}
    m_{\mathrm{opt}}
    =
        \left\lceil
        \frac{\pi}{4 \theta_w}
        -
        \frac{1}{2}
        \right\rceil
    .
    \label{gen_loc_state:m_opt_for_QAA}
\end{align}
As explained above,
we can reduce $w$ continuously to $w_{\mathrm{AR}}$ via the amplitude reduction.
We can thus make $\theta_{w_{\mathrm{AR}}}$ coincide with $\theta^{(m_{\mathrm{opt}}, 0)}$
by setting the angle parameter for amplitude reduction in
Eq.~(\ref{gen_loc_state:success_prob_after_ampl_reduction})
as
\begin{align}
    \theta_{\mathrm{AR,opt}}
    =
        \arccos
        \left(
            \frac{1}{\sqrt{w}}
            \sin \frac{\pi}{4 m_{\mathrm{opt}} + 2}
        \right)
    .
    \label{gen_loc_state:angle_for_determinization}
\end{align}
These observations indicate that
$\mathcal{C}_{\mathrm{prob}}^{(\mathrm{AR})}$ with  $\theta_{\mathrm{AR,opt}}$ and the subsequent QAA render the nonunitary operation of $\mathcal{A}$ deterministic regardless of $w.$
We refer to this treatment as the determinization of the probabilistic operation and denote it by the quantum amplitude reduction and amplification (QARA) technique.
While the expected operation number for the determinized circuit does not differ largely than for the probabilistic circuit,
the striking feature of the former is that the omission of measurement is justified at the algorithmic level.
This helps to develop an efficient algorithm without paying attention to measurement overhead,
readout errors, and reset of ancillae.
Since $w$ is on the order of $1/n_{\mathrm{loc}},$
we have $m_{\mathrm{opt}} = \mathcal{O} (\sqrt{n_{\mathrm{loc}}})$ from
Eq.~(\ref{gen_loc_state:m_opt_for_QAA}).
In particular, for a case where $1/4 \leq w < 1,$
we have $m_{\mathrm{opt}} = 1.$ 
Nishi et al.~\cite{Nishi_QAA} have proposed a similar technique for determinizing the QAA for probabilistic imaginary-time evolution (PITE)\cite{bib:5737, bib:6103, bib:6231, bib:6242} steps.
The quadratic speedup of PITE \cite{bib:6571} with determinization will be a good scheme for nonvariational energy minimization on fault-tolerant quantum computers.

\subsubsection{Error analysis for inaccurate QAE}

Although we have adopted QARA for the determinization,
the weight $w$ of a success state can also be raised to unity by employing the technique in the original paper \cite{bib:4878} of QAA,
which modifies the reflection circuit at the final iteration.
We denote this technique by the modified QAA.
Thereby a complex transcendental equation has to be solved using $w$ obtained via QAE to decide the circuit parameters,
and the dependence of the error in the amplified state on that in the QAE process is thus difficult to evaluate.
Also,
the fixed-point search (FPS) technique \cite{bib:6184} amplifies the unknown weight $w$ of a success state to $1 - \delta^2$ for a specified tolerance $\delta.$
$\delta$ must be strictly larger than zero for achieving the quadratic speedup.
The modified QAA and FPS techniques do not require an additional qubit in contrast to QARA.

Let us consider a practical situation
where the observer obtains an erroneous estimation $w^{(\mathrm{est})} = w + \Delta w$ of the weight $w$ of the original success state.
We do not introduce any assumption about where the error $\Delta w$ has come from.
It is noted here that a statistical error can intrude into the estimation even when the quantum computation is performed fault-tolerantly due to a finite number of measurements.
The circuit parameter $\theta_{\mathrm{AR, opt}}$ for amplitude reduction and the repetition number $m_{\mathrm{opt}}$ for amplitude amplification can now involve errors.
We demonstrate in
Appendix \ref{sec:error_analysis_for_ampl_red}
that QARA allows us to derive a simple relation between the error in the QAE process and that in the determinization.
When we approach the limit of $w \to 0$ while keeping $\Delta w/w$ finite,
an approximate maximum of the weight of the failure state after the completion of the QARA process is given by
\begin{align}
    \epsilon_{\mathrm{QARA}}
    =
        \sin^2 \frac{\pi \Delta w}{4 w}
        .
    \label{error_max_of_QARA}
\end{align}
Figure \ref{fig:lin_combo_of_unitaries_and_qara}(c) shows the weight $W_{\mathrm{f}}$ of the failure state after the QARA process for
$\Delta w/w = 0.1, 0.04,$ and $0.01$ as a function of $w.$
$W_{\mathrm{f}}$ after the QAA-only process (without amplitude reduction) is also shown in the figure.
While the values of $W_{\mathrm{f}}$ from QAA-only are seen to scatter for each of $\Delta w/w,$
those from QARA exhibit much smaller variations over the wide range of $w.$
Furthermore,
$\epsilon_{\mathrm{QARA}}$ approximates $W_{\mathrm{f}}$ well.
These observations suggest that QARA is an error-controllable method for determinizing a probabilistic operation.

\subsection{Probabilistic encoding of an LC of localized functions}
\label{sec:prob_encoding_of_LC_of_loc_funcs}

Here we describe the protocol for encoding an LC of generic localized functions on qubits.

\subsubsection{Setup}

Let us consider a case where
an $n_q$-qubit data register is available and
we are provided with
$n_{\mathrm{loc}}$ real functions $\{ f_\ell \}_{\ell = 0}^{n_{\mathrm{loc}} - 1}$ localized at the origin as an expansion basis set in one-dimensional (1D) space.
For the equidistant grid points $x_j \equiv j \Delta x \ (j = 0, \dots, 2^{n_q} - 1)$ of a spacing $\Delta x$ on a range $[0, N \Delta x],$
we want to encode a normalized LC of the displaced basis functions on the data register as
\begin{align}
    | \psi_{\mathrm{lc}} \rangle
    =
        \sum_{j = 0}^{N - 1}
        \sum_{\ell = 0}^{n_{\mathrm{loc}} - 1}
            d_\ell
            f_\ell (x_j - k_{\mathrm{c} \ell} \Delta x)
            | j \rangle_{n_q}
    ,
    \label{gen_loc_state:state_for_localized_funcs}
\end{align}
where $N \equiv 2^{n_q}.$
$| j \rangle_{n_q}$ is the computational basis for the data register.
$k_{\mathrm{c} \ell}$ is the integer coordinate
$(0 \leq k_{\mathrm{c} \ell} \leq N - 1)$
of the center of the displaced $\ell$th basis function,
where the subscript c is for `center'.
$d_\ell$ is the real coefficient for the LC.
We assume that the basis functions are normalized over the range,
that is, $\sum_{j = 0}^{N - 1} f_{\ell} (x_j)^2 = 1$ for each $\ell,$
and the displaced ones are not necessarily orthogonal to each other.
We further assume that the $n_q$-qubit unitary oracle for generating each basis function centered at the origin is known:
$
U_{\mathrm{orig}}^{(\ell)} | 0 \rangle_{n_q}
=
\sum_{j = 0}^{N - 1}
f_\ell (j \Delta x)
| j \rangle_{n_q}
$
for each $\ell.$
The assumption for the center to be the origin is in fact not necessary
since we can cancel the discrepancy by adjusting the parameters in the phase shift unitaries introduced below.
We adopt the definition of oracle $U_{\mathrm{orig}}^{(\ell)},$
however, for the discussion to be simple.

\subsubsection{Displaced single basis function}

\begin{figure*}
\begin{center}
\includegraphics[width=16cm]{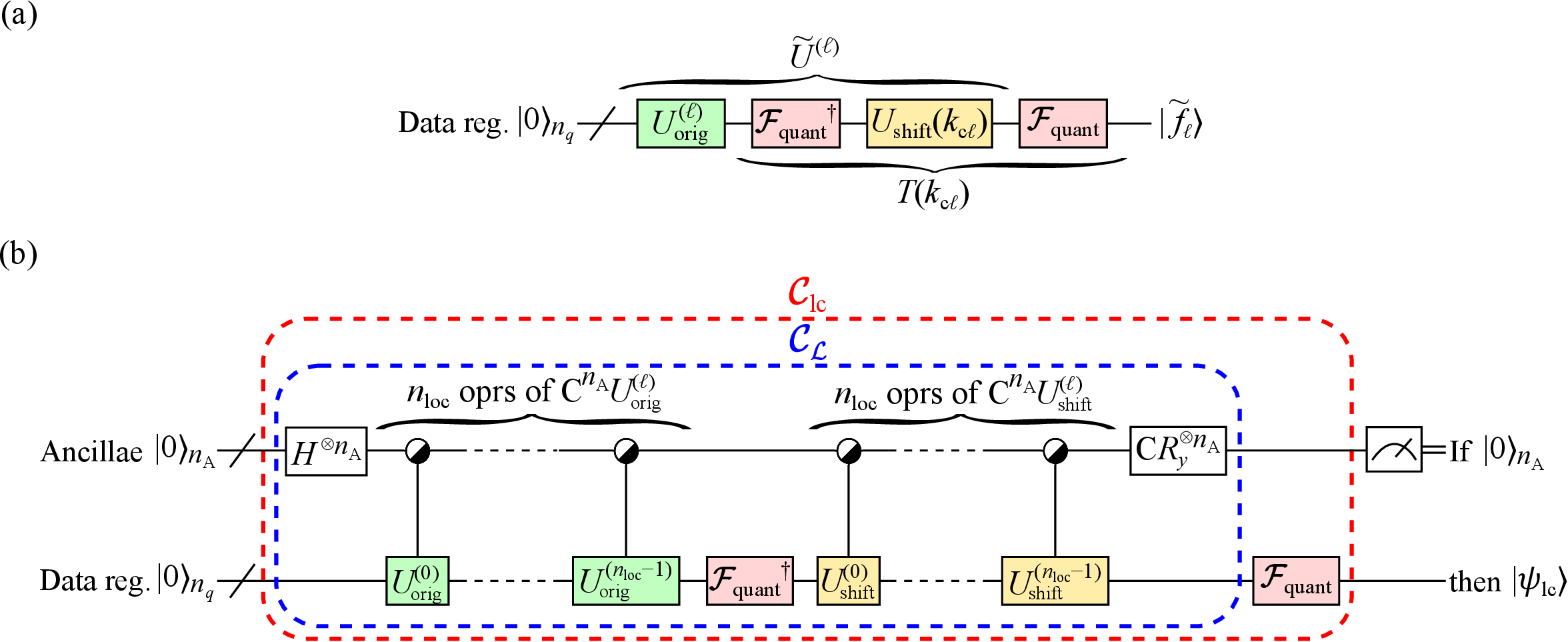}
\end{center}
\caption{
(a)
Circuit that generates the single displaced basis function $| \widetilde{f}_\ell \rangle.$
(b)
Circuit $\mathcal{C}_{\mathrm{lc}}$ that generates probabilistically the LC $| \psi_{\mathrm{lc}} \rangle$ of the displaced basis functions.
The partial circuit $\mathcal{C}_{\mathcal{L}}$ is a special case of $\mathcal{C}_{\mathrm{prob}}$ in
Fig.~\ref{fig:lin_combo_of_unitaries_and_qara}(a).
Each of the half-filled circles in this figure represents the multiple control and anti-control bits.
$\mathrm{C} R_y^{\otimes n_{\mathrm{A}}}$ represents
the single-qubit $y$ rotations on the ancillae controlled by themselves.
For details, see the original paper \cite{bib:5163} for
the implementation of a generic LC of unitaries.
}
\label{fig:circuit_lc_loc_funcs}
\end{figure*}

Let us find the circuit that generates the displaced basis function
\begin{align}
    | \widetilde{f}_\ell \rangle
    \equiv
        \sum_{j = 0}^{N - 1}
            f_\ell (j' \Delta x)
        | j \rangle_{n_q}
        \Big|_{j' = j - k_{\mathrm{c} \ell} \ \mathrm{mod} \ N }
\end{align}
for each $\ell.$
To this end,
we define the phase shift unitary $U_{\mathrm{shift}} (k)$ for an integer $k$ that acts on a computational basis diagonally as
\begin{align}
    U_{\mathrm{shift}} (k)
    | j \rangle_{n_q}
    =
        \exp
        \left(
            -i
            \frac{2 \pi k}{N}
            j
        \right)
        | j \rangle_{n_q}
    .
    \label{gen_loc_state:def_U_shift}
\end{align}
This operation can be implemented as separate single-qubit unitaries,
as explained in Appendix \ref{appendix:impl_of_phase_shift}.
With this and the quantum Fourier transform (QFT) operation $\mathcal{F}_{\mathrm{quant}}$ \cite{Nielsen_and_Chuang}, the operator 
\begin{gather}
    T (k)
    \equiv
        \mathcal{F}_{\mathrm{quant}}
        \cdot
        U_{\mathrm{shift}} (k)
        \cdot
        \mathcal{F}_{\mathrm{quant}}^\dagger
    \label{gen_loc_state:def_translation_opr}
\end{gather}
is easily confirmed to perform modular addition for a computational basis as
$
T (k) | j \rangle_{n_q}
=
| (j + k) \ \mathrm{mod} \ N \rangle_{n_q}
.
$
The displaced basis function can thus be generated by locating the function at the origin and translating it by
$
k_{\mathrm{c} \ell}:
| \widetilde{f}_\ell \rangle
=
T (k_{\mathrm{c} \ell})
U_{\mathrm{orig}}^{(\ell)}
| 0 \rangle_{n_q},
$
as depicted in Fig.~\ref{fig:circuit_lc_loc_funcs}(a).

\subsubsection{LC of displaced basis functions}

We define
\begin{align}
    \widetilde{U}^{(\ell)}
    \equiv
        U_{\mathrm{shift}}^{(\ell)}
        \cdot
        \mathcal{F}_{\mathrm{quant}}^\dagger
        \cdot
        U_{\mathrm{orig}}^{(\ell)}
    \label{gen_loc_state:gen_shifted_loc_func}
\end{align}
with
$
U_{\mathrm{shift}}^{(\ell)}
\equiv
U_{\mathrm{shift}} (k_{\mathrm{c} \ell})
$
for each $\ell.$
With the LC
$
    \mathcal{L}
    \equiv
        \sum_{\ell = 0}^{n_{\mathrm{loc}} - 1}
            d_\ell
        \widetilde{U}^{(\ell)}
$
of them,
the desired state in 
Eq.~(\ref{gen_loc_state:state_for_localized_funcs})
is written as
$
    | \psi_{\mathrm{lc}} \rangle
    =
        \sum_{\ell = 0}^{n_{\mathrm{loc}} - 1}
            d_\ell
            | \widetilde{f}_\ell \rangle
    =
        \mathcal{F}_{\mathrm{quant}}
        \cdot
        \mathcal{L}
        | 0 \rangle_{n_q}
    .
$
$\mathcal{L},$ which is nonunitary in general,
admits the probabilistic implementation as a special case of $\mathcal{C}_{\mathrm{prob}}$ in 
Fig.~\ref{fig:lin_combo_of_unitaries_and_qara}(a).
Specifically,
the circuit $\mathcal{C}_{\mathrm{lc}}$ in
Fig.~\ref{fig:circuit_lc_loc_funcs}(b)
for the data register and 
\begin{align}
    n_{\mathrm{A}}
    \equiv
        \lceil
        \log_2 n_{\mathrm{loc}}
        \rceil
    \label{gen_loc_state:def_num_of_ancillae}
\end{align}
ancillae encodes $| \psi_{\mathrm{lc}} \rangle$ probabilistically.
The central part, $\mathcal{C}_{\mathcal{L}},$
implements $\mathcal{L}$ probabilistically by
recognizing the $n_{\mathrm{A}}$-bit integer $\ell$ specified by an ancillary state $| \ell \rangle_{n_{\mathrm{A}}}.$
For details, see the original paper \cite{bib:5163},
which describes the implementation of a generic LC of unitaries.
It is noted here that a slight modification to the circuit in the original paper exists in $\mathcal{C}_{\mathcal{L}}.$
That is, since 
$\mathcal{F}_{\mathrm{quant}}^\dagger$ in $\widetilde{U}^{(\ell)}$ does not refer to $\ell$
[see Eq.~(\ref{gen_loc_state:gen_shifted_loc_func})],
all the controlled $\mathcal{F}_{\mathrm{quant}}^\dagger$ have been replaced by the uncontrolled $\mathcal{F}_{\mathrm{quant}}^\dagger,$
as seen in Fig.~\ref{fig:circuit_lc_loc_funcs}(b).
Since the $\mathcal{F}_{\mathrm{quant}}$ operation after $\mathcal{C}_{\mathcal{L}}$ does not act on the ancillae,
it is also possible to perform the measurement,
observe the success state,
and then operate $\mathcal{F}_{\mathrm{quant}}$ on the data register.

Let us estimate the depth of $\mathcal{C}_{\mathrm{lc}}$ by assuming that of $U_{\mathrm{orig}}^{(\ell)}$ is independent of $\ell.$
Since
$\mathrm{depth}(\mathcal{F}_{\mathrm{quant}}) = \mathcal{O} (n_q)$
\cite{bib:6590,bib:6591}
and
$\mathrm{depth} (U_{\mathrm{shift}}^{(\ell)}) = \mathcal{O} (1)$
for each $\ell,$ we find
\begin{align}
    \mathrm{depth} 
    \left(
        \mathcal{C}_{\mathrm{lc}}
    \right)
    &=
        \mathcal{O}
        \left(
            \max
            \left(
                n_{\mathrm{loc}}
                \mathrm{depth}
                \left(
                    \mathrm{C}^{n_{\mathrm{A}}}
                    U_{\mathrm{orig}}^{(\ell)}
                \right)
                ,
                n_q
            \right)
        \right)
    .
    \label{gen_loc_state:whole_depth_lin_combo_naive_impl}
\end{align}

Although we have assumed the basis functions and the coefficients to be real,
it is possible to encode any LC of complex basis functions via complex coefficients by redefining the unitaries and coefficients, as explained in
Appendix \ref{appendix:complex_funcs_and_complex_coeffs}.
The descriptions below for real values can thus apply to complex values with small modifications for $n_{\mathrm{A}} + 1$ ancillae.
Also, the protocol described above can be extended to multidimensional space by employing product basis functions,
as explained in
Appendix \ref{appendix:multidimensional_space}.

\subsection{Discrete Slater and Lorentzian functions}
\label{sec:discrete_Slater_and_Lorentzian_funcs}

\subsubsection{Discrete Slater functions}

For establishing the protocol for encoding an LC of LFs,
we first introduce discrete Slater functions (SFs)
since they are related to the corresponding LFs via QFT, as will be demonstrated later.

Klco and Savage \cite{bib:5645} have proposed a circuit that encodes a symmetric exponential function,
based on which we design the discrete SFs and the circuit construction for them.
Specifically,
for an $n_q$-qubit system with $N \equiv 2^{n_q},$
we define a discrete SF $S (n_q, a)$ having a decay rate $a > 0$ such that its value at an integer coordinate $j$ is
\begin{align}
    S_j (n_q, a)
    \equiv
        \begin{cases}
            C_{S} (n_q, a) e^{-a \widetilde{j}}
            &
            0 \leq \widetilde{j} < N/2 \\
            C_{S} (n_q, a) e^{-a (N - \widetilde{j})}
            &
            N/2 \leq \widetilde{j} < N
        \end{cases}
    \label{gen_loc_state:def_discrete_Slater}
\end{align}
with the normalization constant
\begin{align}
    C_{S} (n_q, a)
    \equiv
        \sqrt{\frac{1 - e^{-2 a}}{(1 + e^{-2 a}) (1 - e^{- N a}) }}
    .
    \label{gen_loc_state:Slater_state_normalization_const}
\end{align}
We have introduced the tilde symbol as
$\widetilde{j} = j \ \mathrm{mod} \ N.$
This tilde symbol is used for other integers similarly in what follows.
$S (n_q, a)$ is a period-$N$ function having cusps at the origin and its duplicated points,
while each of the peaks in the encoded function by Klco and Savage \cite{bib:5645} is characterized by neighboring two points that take the same value.

\begin{figure*}
\begin{center}
\includegraphics[width=17cm]{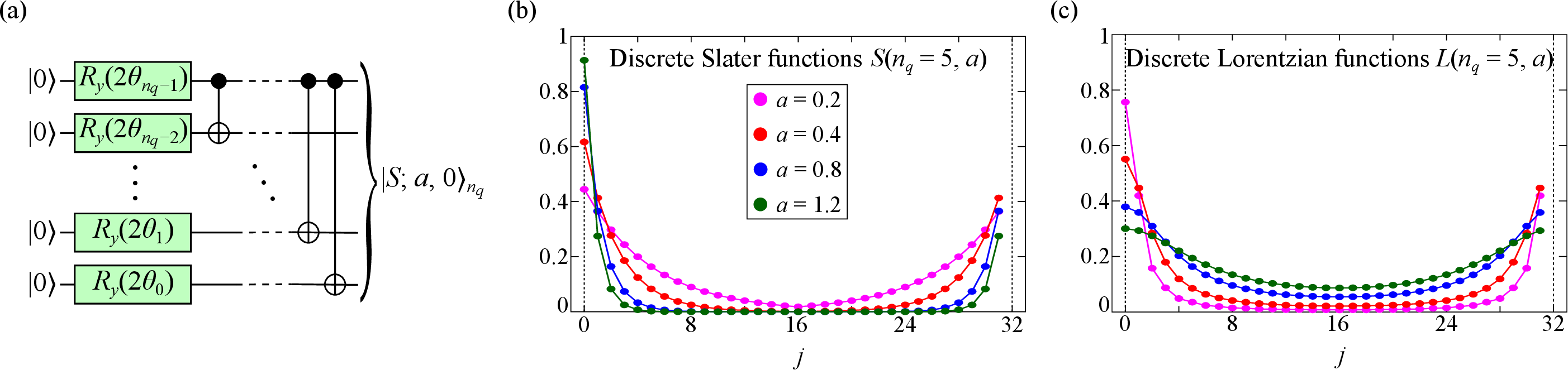}
\end{center}
\caption{
(a)
Circuit $U^{(\mathrm{S})}$ that generates the SF state $| S; a, 0 \rangle_{n_q}.$
(b)
$S (n_q = 5, a)$ for some values of decay rates $a.$
The functions are plotted for integer coordinates $0 \leq j \leq N - 1.$ 
(c)
$L (n_q = 5, a)$ for some values of decay rates $a.$
The correspondence between the decay rates and the colors in this plot is the same as in (b).
}
\label{fig:Slater_and_Lorentzian}
\end{figure*}

We define the normalized SF state centered at an integer coordinate $k_{\mathrm{c}}$ as
\begin{align}
    | S ; a, k_{\mathrm{c}} \rangle_{n_q}
    \equiv
        \sum_{j = 0}^{N - 1}
        S_{j - k_{\mathrm{c}} } (n_q, a)
        | j \rangle_{n_q}
    .
    \label{gen_loc_state:def_Slater_func_state}
\end{align}
$| S ; a, 0 \rangle_{n_q}$ can be generated from initialized qubits by constructing the circuit $U^{(\mathrm{S})}$ as shown in
Fig.~\ref{fig:Slater_and_Lorentzian}(a),
where the angle parameters for the $y$ rotation on $| q_m \rangle$ is given by
\begin{align}
    \theta_m
    \equiv
    \begin{cases}
        \arctan e^{-a} & m = n_q - 1 \\
        \arctan \exp (-2^m a) & \mathrm{Otherwise}
    \end{cases}
    .
    \label{gen_loc_state:def_rot_angle_for_exp_state}
\end{align}
For details, see Appendix \ref{appendix:action_of_U_S}.
We adopt the logarithmic-depth implementation for CNOT gates (see Appendix \ref{sec:impl_for_multi_ctrl_X_gates}) to achieve the entire depth
\begin{align}
    \mathrm{depth}
    ( U^{(\mathrm{S})} )
    =
        \mathcal{O} (\log n_q)
    .
    \label{circuit:gen_loc_state:depth_Slater_at_orig}
\end{align}

The discrete SFs for five data qubits having various decay rates are plotted in 
Fig.~\ref{fig:Slater_and_Lorentzian}(b) as examples.

\subsubsection{Discrete Lorentzian functions}

It is well known that continuous SFs and LFs are related with each other via the Fourier transform.
Recalling this fact,
we define a discrete LF $L (n_q, a)$ such that its value at an integer coordinate $j$ is
\begin{align}
    L_j (n_q, a)
    \equiv
        \frac{C_{S} (n_q, a) }{\sqrt{N}}
        \frac{(1 - e^{-2 a}) (1 - (-1)^j e^{-a N/2})}
        { 1 - 2 e^{-a} \cos (2 \pi j/N) + e^{-2 a} }
        .
        \label{gen_loc_state:def_discrete_Lorentzian}
\end{align}
$a$ is a real positive parameter.
This period-$N$ function has peaks at the origin and its duplicated points, as well as the discrete SF.
The discrete LFs for five data qubits having various decay rates are plotted in 
Fig.~\ref{fig:Slater_and_Lorentzian}(c) as examples.

We define the normalized LF state centered at an integer coordinate $k_{\mathrm{c}}$ as
\begin{align}
    | L ; a, k_{\mathrm{c}} \rangle_{n_q}
    \equiv
        \sum_{j = 0}^{N - 1}
        L_{j - k_{\mathrm{c}}} (n_q, a)
        | j \rangle_{n_q}
    .
    \label{gen_loc_state:def_Lorentzian_func_state}
\end{align}
$| L; a, 0 \rangle_{n_q}$ is symmetric around $j = N/2.$
As demonstrated in Appendix \ref{appendix:Lorentzian_from_Slater},
the SF and LF states with a common decay rate at the origin are related with each other via QFT:
$
| L ; a, 0 \rangle_{n_q}
=
\mathcal{F}_{\mathrm{quant}}
| S ; a, 0 \rangle_{n_q}
.
$
This means that the unitary
\begin{align}
    U^{(\mathrm{L})}
    \equiv
        \mathcal{F}_{\mathrm{quant}}
        \cdot
        U^{(\mathrm{S})}
    \label{gen_loc_state:Lorentzian_from_Slater}
\end{align}
generates the LF state from initialized qubits.
$\mathcal{F}_{\mathrm{quant}}$ on the right-hand side of
Eq.~(\ref{gen_loc_state:Lorentzian_from_Slater})
can be replaced by $\mathcal{F}_{\mathrm{quant}}^\dagger.$
As seen in Figs.~\ref{fig:Slater_and_Lorentzian}(b) and (c),
a discrete SF having a larger decay rate corresponds to
a discrete LF having a wider peak.
In fact,
from Eq.~(\ref{gen_loc_state:def_discrete_Lorentzian}),
the discrete LF behaves near the origin
$(| j | \ll N)$ approximately as
$L_j (n_q, a) \propto 1/(j^2 + \gamma_L (a)^2),$
where $ \gamma_{L} (a) \equiv (N/\pi) \sinh (a/2)$
is a monotonically increasing function of $a.$
In what follows, we denote $a$ for $L (n_q, a)$ simply as the decay rate of the discrete LF since no confusion would occur.
The overlap between two LFs is given as a simple expression, as derived in
Appendix \ref{appendix:overlap_btwn_Lorentzian_states}.
The circuit depth of $U^{(\mathrm{L})}$
implemented as
Eq.~(\ref{gen_loc_state:Lorentzian_from_Slater}) is,
from Eq.~(\ref{circuit:gen_loc_state:depth_Slater_at_orig}),
\begin{align}
    \mathrm{depth} ( U^{(\mathrm{L})} )
    =
        \mathcal{O} (n_q)
    .
    \label{circuit:gen_loc_state:depth_Lorentz_at_orig}
\end{align}

\subsection{Implementation for an LC of discrete Lorentzian functions}

\subsubsection{Probabilistic encoding}

We know the unitaries $U^{(\mathrm{S})}$ and $U^{(\mathrm{L})}$ for generating an SF and an LF at the origin, respectively,
as described in 
Sect.~\ref{sec:discrete_Slater_and_Lorentzian_funcs}.
We can thus construct explicitly the circuit for encoding probabilistically an LC $| \psi_{\mathrm{lc}} \rangle$ of SFs and/or LFs according to 
Sect.~\ref{sec:prob_encoding_of_LC_of_loc_funcs}.
Although the circuit of $U^{(\mathrm{S})}$ is shallower than that of $U^{(\mathrm{L})}$
[see Eqs.~(\ref{circuit:gen_loc_state:depth_Slater_at_orig}) and (\ref{circuit:gen_loc_state:depth_Lorentz_at_orig})],
the circuits for displaced single SF and LF have the common scaling $\mathcal{O} (n_q)$ due to the QFT operations,
as found in Fig.~\ref{fig:circuit_lc_loc_funcs}(a).
When we adopt the LFs as the basis functions for the expansion in 
Eq.~(\ref{gen_loc_state:state_for_localized_funcs}),
$\mathcal{F}_{\mathrm{quant}}^\dagger$ in
Fig.~\ref{fig:circuit_lc_loc_funcs}(a)
cancels
$\mathcal{F}_{\mathrm{quant}}$ in $U^{(\mathrm{L})},$ as understood from
Eq.~(\ref{gen_loc_state:Lorentzian_from_Slater}).
This means that we can generate the displaced single LF by implementing $U^{(\mathrm{S})}$
and performing $\mathcal{F}_{\mathrm{quant}}$ only once.
This favorable property of the LFs is inherited by the circuit $\mathcal{C}_{\mathrm{lc}}$ in
Fig.~\ref{fig:circuit_lc_loc_funcs}(b)
for the desired LC.
We therefore adopt the LFs as the basis set in what follows:
\begin{align}
    | \psi_{\mathrm{lc}} \rangle
    =
        \sum_{\ell = 0}^{n_{\mathrm{loc}} - 1}
            d_\ell
            | L; a_\ell, k_{\mathrm{c} \ell} \rangle_{n_q}
    .
    \label{gen_loc_state:LC_of_LFs}
\end{align}
The normalization condition for the expansion coefficients are provided explicitly in
Appendix \ref{appendix:norm_cond_for_an_LC_of_LFs}.

\begin{figure*}
\begin{center}
\includegraphics[width=16cm]{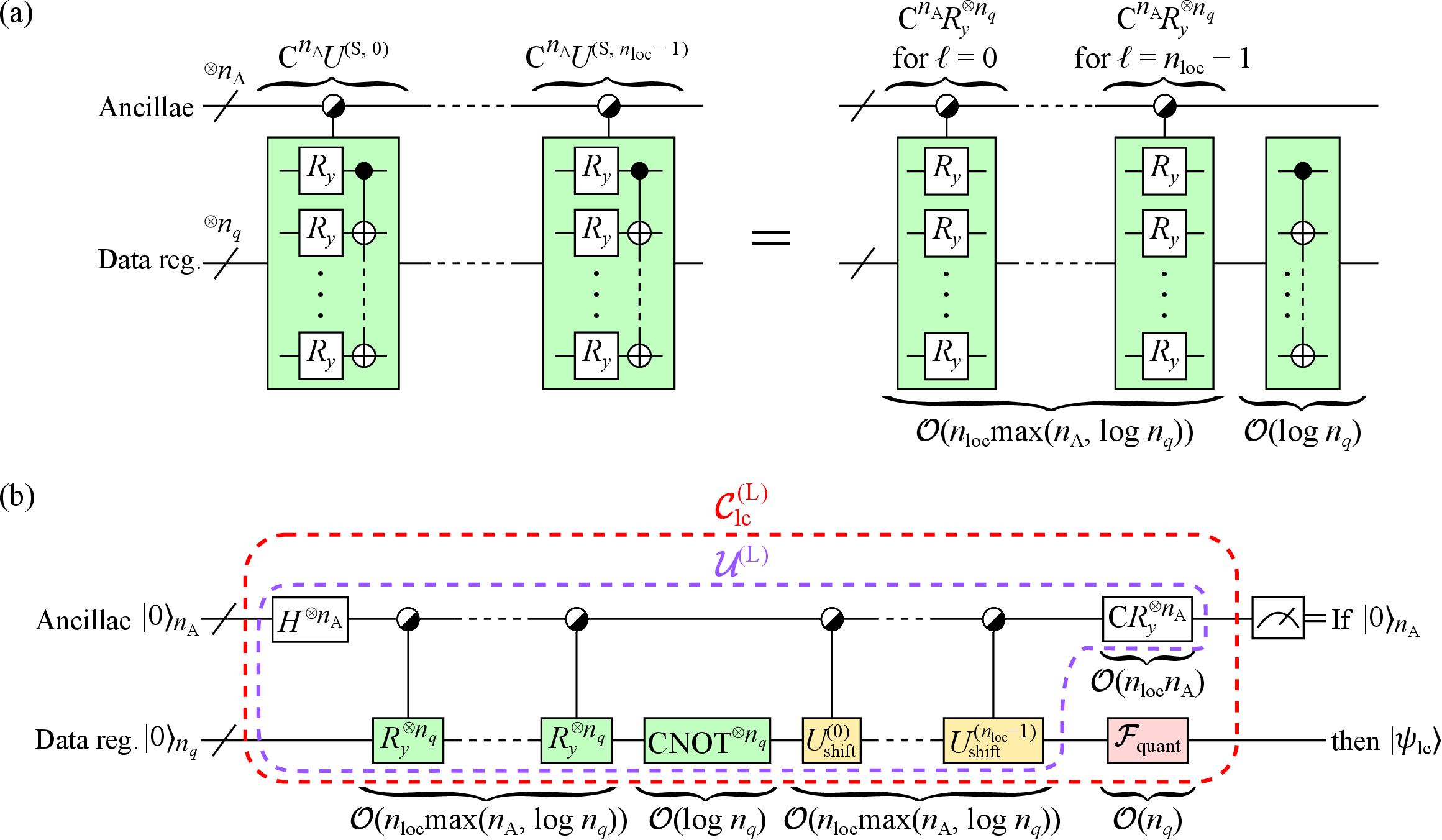}
\end{center}
\caption{
(a)
Left circuit is a naive implementation of
the controlled $U^{(\mathrm{S})}$ unitaries for the $n_{\mathrm{loc}}$ LFs at the origin.
The right circuit, equivalent to the left one,
no longer requires the ancillary control bits for the CNOTs.
The scaling of circuit depth with respect to $n_{\mathrm{loc}}, n_{\mathrm{A}},$ and $n_q$ are also shown in this figure.
(b)
Circuit $\mathcal{C}_{\mathrm{lc}}^{(\mathrm{L})}$
for probabilistic encoding of the LC of LFs as a special case of $\mathcal{C}_{\mathrm{lc}}$ in
Fig.~\ref{fig:circuit_lc_loc_funcs}(b).
}
\label{fig:circuit_lc_Lorentzian_prob}
\end{figure*}

We denote the $U^{(\mathrm{S})}$ for the $\ell$th basis function by $U^{(\mathrm{S}, \ell)}.$
Since it is constructed as shown in 
Fig.~\ref{fig:Slater_and_Lorentzian}(a),
a naive implementation of $\mathcal{C}_{\mathrm{lc}}$ requires
a series of CNOT gates for each $\ell$ controlled by the ancillae,
as depicted as the left circuit in
Fig.~\ref{fig:circuit_lc_Lorentzian_prob}(a).
The consecutive CNOTs act on the data register only once for each $\ell.$
The control bits for these CNOTs can thus be removed,
leading to the right circuit in the figure.
The circuit $\mathcal{C}_{\mathrm{lc}}^{(\mathrm{L})}$
as a special case of $\mathcal{C}_{\mathrm{lc}}$
is then constructed as shown in
Fig.~\ref{fig:circuit_lc_Lorentzian_prob}(b).
The right circuit in
Fig.~\ref{fig:circuit_lc_Lorentzian_prob}(a)
contains the sequence of
multiply controlled multiple single-qubit (MCM1) unitaries.
The partial circuit for each $\ell$ can be implemented by using the technique described in
Appendix \ref{sec:impl_for_multi_ctrl_multi_1q},
leading to the depth of
$\mathcal{O} (\max (n_{\mathrm{A}}, \log n_q)).$

Let us estimate the scaling of depth of $\mathcal{C}_{\mathrm{lc}}^{(\mathrm{L})}$ with respect to $n_q, n_{\mathrm{loc}},$ and $n_{\mathrm{A}}.$
The implementation technique for MCM1 unitaries is also applicable to the part for the phase shift unitaries.
The controlled $y$ rotations among the ancillae can be implemented with a depth of 
$\mathcal{O} (n_{\mathrm{loc}} n_{\mathrm{A}})$ \cite{bib:5614, bib:5662, bib:5995}.
The total depth is thus estimated to be
\begin{align}
    \mathrm{depth} 
    \left(
        \mathcal{C}_{\mathrm{lc}}^{(\mathrm{L})}
    \right)
    &=
        \mathcal{O}
        \left(
            \max
            \left(
                n_{\mathrm{loc}}
                \log n_{\mathrm{loc}}
                ,
                n_{\mathrm{loc}}
                \log n_q
                ,
                n_q 
            \right)
        \right)
    ,
    \label{gen_loc_state:whole_depth_lin_combo_using_Lorentzians}
\end{align}
where we used
Eq.~(\ref{gen_loc_state:def_num_of_ancillae}).
As stated in Sect.~\ref{sec:generic_determinization},
the success probability is on the order of
$1/2^{n_{\mathrm{A}}} \approx 1/n_{\mathrm{loc}}.$
Since the operation of QFT on the data register can be delayed until the success state is observed,
the computational time spent until the encoding is done is estimated to be
\begin{align}
    \mathrm{time}
    \left(
        \mathcal{C}_{\mathrm{lc}}^{(\mathrm{L})}
    \right)
    &=
        \mathcal{O}
        \left(
            n_{\mathrm{loc}}
            \mathrm{depth} 
            (\mathcal{U}^{(\mathrm{L})})
            +
            \mathrm{depth} (\mathcal{F}_{\mathrm{quant}})
        \right)
    \nonumber \\
    &=
        \mathcal{O}
        \left(
            \max
            \left(
                n_{\mathrm{loc}}^2
                \log n_{\mathrm{loc}}
                ,
                n_{\mathrm{loc}}^2
                \log n_q
                ,
                n_q 
            \right)
        \right)
        .
    \label{gen_loc_state:runtime_Lorentzian}
\end{align}

\subsubsection{Deterministic encoding}

\begin{figure*}
\begin{center}
\includegraphics[width=17cm]{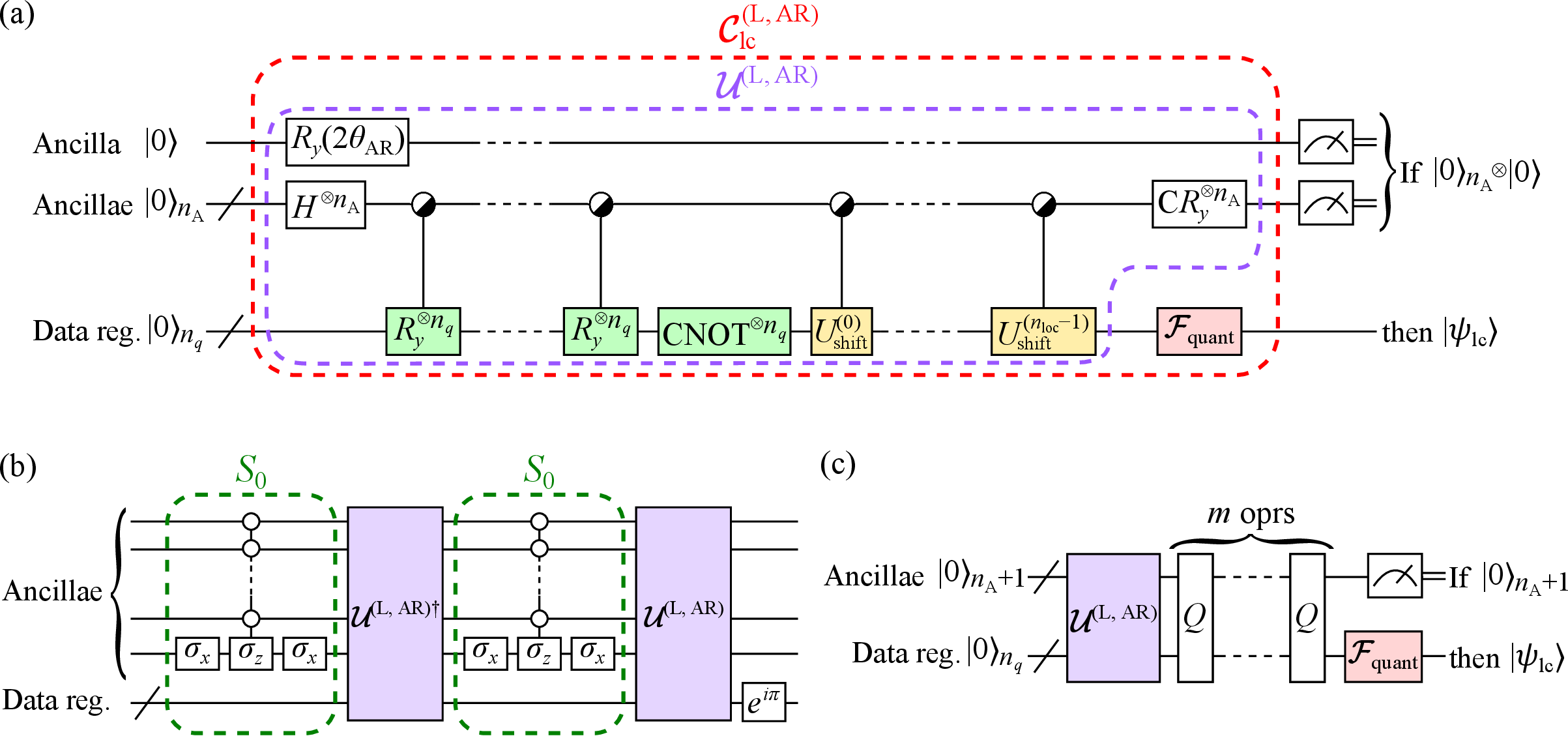}
\end{center}
\caption{
(a)
Circuit $\mathcal{C}_{\mathrm{lc}}^{(\mathrm{L, AR})}$
equipped with an extra ancilla for amplitude reduction
for probabilistic encoding of the LC $| \psi_{\mathrm{lc}} \rangle$ of discrete LFs.
(b)
Amplification operator $Q$ for QAA.
$S_0$ represents the zero reflection operation defined in Eq.~(\ref{def_zero_reflection}).
(c)
Circuit $\mathcal{C}_{\mathrm{lc}}^{(\mathrm{L}, m)}$
where QAA has been applied via $Q^m.$
The encoding using this circuit 
with a generic combination of $m$ and $\theta_{\mathrm{AR}}$
is still probabilistic.
When $m_{\mathrm{opt}}$ and $\theta_{\mathrm{AR,opt}}$ are adopted,
the circuit $\mathcal{C}_{\mathrm{lc}}^{(\mathrm{L,det})}$
achieves deterministic encoding.
}
\label{fig:circuit_lc_Lorentzian_det}
\end{figure*}

The encoding using $\mathcal{C}_{\mathrm{lc}}^{(\mathrm{L})}$ is probabilistic, as explained just above.
Let us determinize it according to the QARA technique.
The amplitude reduction is achieved by constructing the circuit
$\mathcal{C}_{\mathrm{lc}}^{(\mathrm{L,AR})}$ in
Fig.~\ref{fig:circuit_lc_Lorentzian_det}(a).
The degree to which the weight of success state is reduced is tuned by the angle parameter $\theta_{\mathrm{AR}}$ for the extra ancilla.

Since the QFT operations in
$\mathcal{C}_{\mathrm{lc}}^{(\mathrm{L,AR})}$
does not act on the $n_{\mathrm{A}} + 1$ ancillae,
the amplification operator $Q$ for QAA \cite{bib:4884, bib:4878} can be constructed from the partial circuit $\mathcal{U}^{(\mathrm{L,AR})}$
[see Fig.~\ref{fig:circuit_lc_Lorentzian_det}(a)] as
$
    Q
    \equiv 
        -
        \mathcal{U}^{(\mathrm{L, AR})}
        S_0
        \mathcal{U}^{(\mathrm{L, AR}) \dagger}
        S_0
    ,
$
where
\begin{align}
    S_0
    \equiv
        I_{\mathrm{d}}
        \otimes
        \left(
            \sigma_x^{\otimes ( n_{\mathrm{A}} + 1 )}
            \cdot
            \mathrm{C}^{n_{\mathrm{A}} } \sigma_z
            \cdot
            \sigma_x^{\otimes ( n_{\mathrm{A}} + 1 )}
        \right)
    \label{def_zero_reflection}
\end{align}
is the so-called zero reflection operator,
that acts nontrivially only when the ancillary state is $| 0 \rangle_{n_{\mathrm{A}} + 1}$ to invert the sign of the state.
$I_{\mathrm{d}}$ is the identity for the data register.
The circuit for $Q$ is shown in
Fig.~\ref{fig:circuit_lc_Lorentzian_det}(b).
For a given number $m$ of operations of $Q,$
QAA is implemented by constructing the circuit $\mathcal{C}_{\mathrm{lc}}^{(\mathrm{L}, m)}$ shown in
Fig.~\ref{fig:circuit_lc_Lorentzian_det}(c).
This circuit performs QFT only once regardless of $m$
thanks to the eliminated QFT operations in $T$ and $U^{(\mathrm{L})}.$

The encoding is still probabilistic for
$\mathcal{C}_{\mathrm{lc}}^{(\mathrm{L}, m)}$ with a generic combination of $m$ and $\theta_{\mathrm{AR}}.$
We can determinize it, however, by choosing an appropriate combination of them.
Specifically, the optimal value $m_{\mathrm{opt}}$ of $m$ is found from the weight $w$ of success state in
$\mathcal{C}_{\mathrm{lc}}^{(\mathrm{L})}$ according to
Eq.~(\ref{gen_loc_state:m_opt_for_QAA}).
With this, we find $\theta_{\mathrm{AR,opt}}$ from
Eq.~(\ref{gen_loc_state:angle_for_determinization})
for
$\mathcal{C}_{\mathrm{lc}}^{(\mathrm{L}, m_{\mathrm{opt}})}$
to obtain the deterministic circuit
$\mathcal{C}_{\mathrm{lc}}^{(\mathrm{L,det} )},$
which is the main result of this study.
We would like to emphasize here that
we are able to evaluate $w$ analytically
without performing QAE in this case
since we know the expressions for the circuit parameters in $\mathcal{C}_{\mathrm{lc}}^{(\mathrm{L})}$ (see Appendix A in Ref.~\cite{bib:5163}).

Let us estimate the scaling of the depth of $\mathcal{C}_{\mathrm{lc}}^{(\mathrm{L}, m)}$ for a given $m.$
From
$
\mathrm{depth} (\mathcal{U}^{(\mathrm{L, AR})})
=
\mathrm{depth} (\mathcal{U}^{(\mathrm{L})})
$
[see Fig.~\ref{fig:circuit_lc_Lorentzian_prob}(b)]
and
$\mathrm{depth} (S_0) = \mathcal{O} (n_{\mathrm{A}})$
\cite{bib:5614, bib:5662, bib:5995},
we find
$
\mathrm{depth} (Q)
= 
\mathcal{O}
( n_{\mathrm{loc}} \max (n_{\mathrm{A}}, \log n_q) ).
$
The total depth thus scales as
\begin{gather}
    \mathrm{depth}
    \left(
        \mathcal{C}_{\mathrm{lc}}^{(\mathrm{L}, m)}
    \right)
    \nonumber \\
    =
        \mathcal{O}
        \left(
            \max
            (
                m
                n_{\mathrm{loc}}
                \log n_{\mathrm{loc}}
                ,
                m
                n_{\mathrm{loc}}
                \log n_q
                ,
                n_q
            )
        \right)
        .
    \label{gen_loc_state:depth_Lorentzian_with_QAA}
\end{gather}
In particular, that of the deterministic circuit scales as
\begin{gather}
    \mathrm{depth}
    \left(
        \mathcal{C}_{\mathrm{lc}}^{(\mathrm{L,det})}
    \right)
    \nonumber \\
    =
        \mathcal{O}
        \left(
            \max
            \left(
                n_{\mathrm{loc}}^{3/2}
                \log n_{\mathrm{loc}}
                ,
                n_{\mathrm{loc}}^{3/2}
                \log n_q
                ,
                n_q
            \right)
        \right)
        \label{gen_loc_state:depth_Lorentzian_deterministic}    
\end{gather}
since $m_{\mathrm{opt}} = \mathcal{O} (\sqrt{n_{\mathrm{loc}}}).$
The scaling of the computational time for
$\mathcal{C}_{\mathrm{lc}}^{(\mathrm{L,det})}$
until success is the same as 
Eq.~(\ref{gen_loc_state:depth_Lorentzian_deterministic})
since the success probability is unity.
This scaling is more favorable than the probabilistic case in 
Eq.~(\ref{gen_loc_state:runtime_Lorentzian}).

\subsection{Resource estimation for quantum chemistry in real space}

One of the intriguing applications of our scheme is the encoding of a one-electron spatial orbital in a molecular system for quantum chemistry in real space \cite{bib:5373, bib:5372, bib:5328, bib:5824, bib:5737, bib:5658, bib:6103, bib:6236, bib:6242, bib:6455, bib:Horiba}.
Before we start a simulation of real-time dynamics or
an energy minimization procedure toward the ground state of a target molecule,
we have to prepare an initial many-electron state.
In the first-quantized formalism for a quantum computer,
explicit antisymmetrization of an initial state is mandatory \cite{bib:4825, bib:5389, bib:Horiba}.
The efficient antisymmetrization scheme proposed by Berry et al.~\cite{bib:5389} assumes that all the occupied MOs have been prepared.
If we adopt such a scheme,
techniques for generating MOs of various shapes have to be established. 
In this subsection, we analyze the computational cost that is required for encoding MOs as LCs of LFs.

\subsubsection{Three-dimensional Lorentzian basis functions}

\begin{figure}
\begin{center}
\includegraphics[width=6cm]{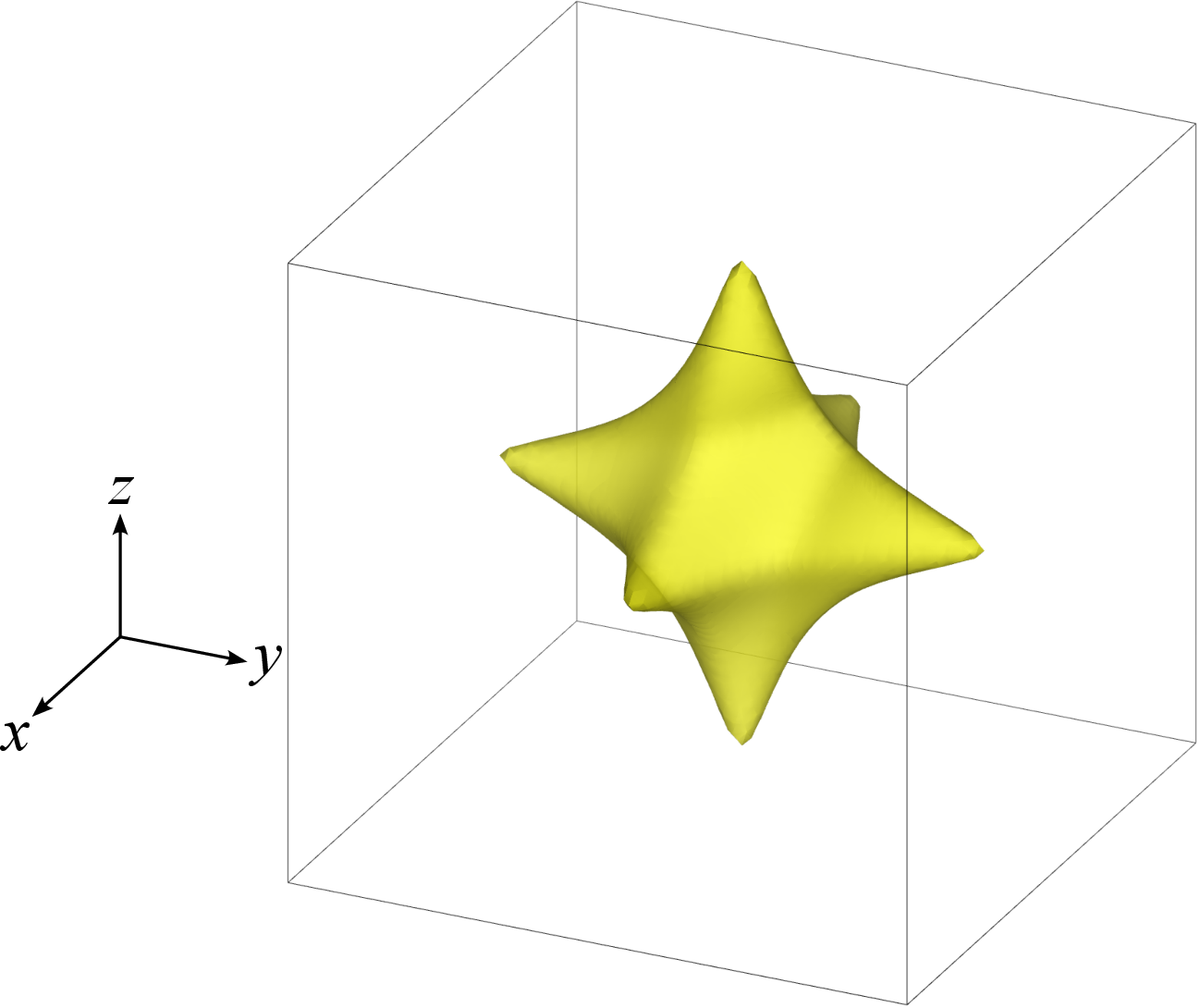}
\end{center}
\caption{
Isosurface of a 3D basis function consisting of a single LF for each direction.
This figure was drawn by using VESTA \cite{VESTA}.
}
\label{fig:Lorentzian_in_3d}
\end{figure}

As explained in Appendix \ref{appendix:multidimensional_space},
our scheme is straightforwardly extended to a three-dimensional (3D) case by employing product basis functions.
The typical shape of a 3D basis function consisting of a single LF for each direction is depicted in Fig.~\ref{fig:Lorentzian_in_3d},
where the shape is a dented octahedron.
Our scheme for a localized spherical function in 3D space leads to an isotropic,
but not rotationally invariant function due to the product basis of the form
$1/(x^2 + a^2) \cdot 1/(y^2 + a^2) \cdot 1/(z^2 + a^2)$
of a width $a.$
That is the price we have to pay for the separability of basis functions.
We can alleviate such rotational asymmetry by increasing the basis functions in each direction around the center.

\subsubsection{Probabilistic encoding}

For a typical target molecule that contains $n_{\mathrm{el}}$ electrons,
the volume of a cubic simulation cell having an edge length $L$ is proportional to the electron number: $n_{\mathrm{el}} \propto L^3.$
The required number of qubits for encoding a one-electron spatial orbital thus scales as
$n_{q e} = \mathcal{O} (\log (n_{\mathrm{el}}^{1/3}/\Delta x))$
for each direction \cite{bib:5737}.
The data register then requires $n_q = 3 n_{q e}$ qubits.

Let us first estimate the computational cost for encoding a generic LC of basis functions,
each of which is a product of three LFs in this case in the simulation cell.
Since the scaling of the depth of probabilistic circuit 
$\mathcal{C}_{\mathrm{lc}}^{(\mathrm{L})}$ in 
Eq.~(\ref{gen_loc_state:whole_depth_lin_combo_using_Lorentzians})
still holds in 3D cases (see Appendix \ref{appendix:multidimensional_space}),
we find
\begin{gather}
    \mathrm{depth} 
    \left(
        \mathcal{C}_{\mathrm{lc}}^{(\mathrm{L})}
    \right)
    \nonumber \\
    =
        \mathcal{O}
        \left(
            \max
            \left(
                n_{\mathrm{loc}}
                \log n_{\mathrm{loc}}
                ,
                n_{\mathrm{loc}}
                \log \log n_{\mathrm{el}}
                ,
                \log n_{\mathrm{el}}
            \right)
        \right)
        .
    \label{gen_loc_state:depth_for_molecule_generic}
\end{gather}

\begin{figure}
\begin{center}
\includegraphics[width=8cm]{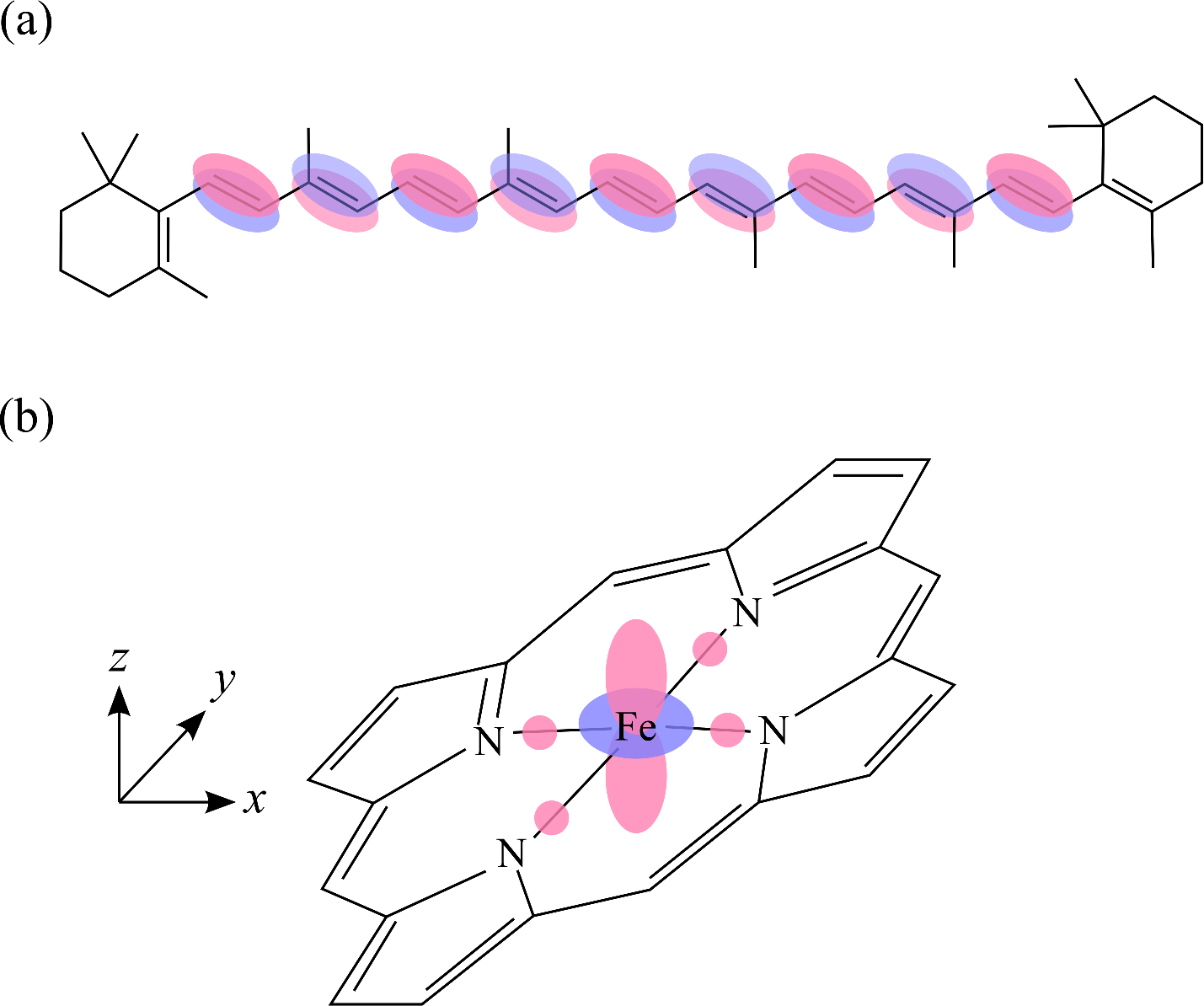}
\end{center}
\caption{
(a)
Sketch of the highest occupied MO of a $\beta$-carotene molecule \cite{bib:6115} as an example of an MO extending over $\mathcal{O} (n_{\mathrm{el}}^1)$ atoms.
The colored lobes designate the regions where the MO has significant amplitudes.
(b)
Sketch of the $d_{z^2}$-derived MO of an iron-porphyrin complex \cite{bib:6118} as an example of an MO extending over $\mathcal{O} (n_{\mathrm{el}}^0)$ atoms.
The porphyrin macrocycle lies on the $xy$ plane.
}
\label{fig:mol_orbs}
\end{figure}

For a case where we want to encode an MO,
which is a chemically motivated LC of one-electron atomic-like orbitals,
we can estimate the computational cost by considering only the electron number. 
Specifically, if the desired MO is delocalized over the molecule
such as the $\pi$ network in a hydrocarbon molecule,
it is made up of the contributions from the constituent atoms with roughly equal weights,
as exemplified in Fig.~\ref{fig:mol_orbs}(a).
The number of basis functions for it thus scales the same way as the molecule size: $n_{\mathrm{loc}} = \mathcal{O} (n_{\mathrm{el}}).$
The circuit depth in this case is, from
Eq.~(\ref{gen_loc_state:depth_for_molecule_generic}),
$
    \mathrm{depth} 
    (
        \mathcal{C}_{\mathrm{lc}}^{(\mathrm{L})}
        \ \mathrm{for \ a \ delocalized \ MO}
    )
    =
        \mathcal{O}
        (
            n_{\mathrm{el}}
            \log n_{\mathrm{el}}
        )
        .
$
If the desired MO is localized, on the other hand,
such as one in a transition metal complex depicted in Fig.~\ref{fig:mol_orbs}(b),
we have clearly $n_{\mathrm{loc}} = \mathcal{O} (1).$
Equation (\ref{gen_loc_state:depth_for_molecule_generic})
in this case reads
$
\mathrm{depth} 
(
\mathcal{C}_{\mathrm{lc}}^{(\mathrm{L})}
\ \mathrm{for \ a \ localized \ MO}
)
=
\mathcal{O} ( \log n_{\mathrm{el}} )
.
$

As for the computational time of the probabilistic encoding,
we find from Eq.~(\ref{gen_loc_state:runtime_Lorentzian}),
\begin{gather}
    \mathrm{time}
    \left(
        \mathcal{C}_{\mathrm{lc}}^{(\mathrm{L})}
    \right)
    \nonumber \\
    =
        \mathcal{O}
        \left(
            \max
            \left(
                n_{\mathrm{loc}}^2
                \log n_{\mathrm{loc}}
                ,
                n_{\mathrm{loc}}^2
                \log \log n_{\mathrm{el}}
                ,
                \log n_{\mathrm{el}}
            \right)
        \right)
        .
    \label{gen_loc_state:time_for_molecule_generic}
\end{gather}
Those for encoding localized and delocalized MOs are estimated as special cases of
Eq.~(\ref{gen_loc_state:time_for_molecule_generic}).
The depth and computational time of
$\mathcal{C}_{\mathrm{lc}}^{(\mathrm{L})}$
are summarized in Table \ref{tab:scaling_mo_using_Lorentzian_with_linear_QFT}.

\begin{table*}[]
\centering
\caption{For encoding an LC of LFs for a molecule in real-space quantum chemistry, this table shows the scaling of depth and computational time in terms of the number $n_{\mathrm{loc}}$ functions and the number $n_{\mathrm{el}}$ of electrons.}
\label{tab:scaling_mo_using_Lorentzian_with_linear_QFT}
\begin{tabular}{ccc}
\hline
MO                   & Depth                                                                                                                                          & Computational time                                                                                                                                \\ \hline
\multicolumn{1}{l}{} & \multicolumn{2}{c}{$\mathcal{C}_{\mathrm{lc}}^{(\mathrm{L})}$ (probabilistic)}                                                                                                                                                                                                                     \\ \cline{2-3} 
Generic              & $\mathcal{O} (  \mathrm{max} ( n_{\mathrm{loc}} \log n_{\mathrm{loc}}, n_{\mathrm{loc}} \log \log n_{\mathrm{el}}, \log n_{\mathrm{el}}   ) )$ & $\mathcal{O} (  \mathrm{max} ( n_{\mathrm{loc}}^2 \log n_{\mathrm{loc}} , n_{\mathrm{loc}}^2 \log \log n_{\mathrm{el}} , \log n_{\mathrm{el}}) )$ \\
Delocalized          & $\mathcal{O} ( n_{\mathrm{el}} \log n_{\mathrm{el}})$                                                                                          & $\mathcal{O} ( n_{\mathrm{el}}^2 \log n_{\mathrm{el}})$                                                                                           \\
Localized            & $\mathcal{O} (  \log n_{\mathrm{el}}  )$                                                                                                       & $\mathcal{O} ( \log n_{\mathrm{el}}  )$                                                                                                           \\
\multicolumn{1}{l}{} & \multicolumn{1}{l}{}                                                                                                                           & \multicolumn{1}{l}{}                                                                                                                              \\
\multicolumn{1}{l}{} & \multicolumn{2}{c}{$\mathcal{C}_{\mathrm{lc}}^{(\mathrm{L,det})}$ (deterministic)}                                                                                                                                                                                                                 \\ \cline{2-3} 
Generic              & \multicolumn{2}{c}{$\mathcal{O} (  \mathrm{max} ( n_{\mathrm{loc}}^{3/2} \log n_{\mathrm{loc}}, n_{\mathrm{loc}}^{3/2} \log \log n_{\mathrm{el}}, \log n_{\mathrm{el}}   ) )$}                                                                                                                     \\
Delocalized          & \multicolumn{2}{c}{$\mathcal{O} ( n_{\mathrm{el}}^{3/2} \log n_{\mathrm{el}} )$}                                                                                                                                                                                                                   \\
Localized            & \multicolumn{2}{c}{$\mathcal{O} ( \log n_{\mathrm{el}}   )$}                                                                                                                                                                                                                                       \\ \hline
\end{tabular}
\end{table*}

\subsubsection{Deterministic encoding}

The computational cost for the deterministic encoding using 
$\mathcal{C}_{\mathrm{lc}}^{(\mathrm{L,det})}$
can also be estimated.
Its depth for generic 1D cases given by
Eq.~(\ref{gen_loc_state:depth_Lorentzian_deterministic})
still holds in 3D cases.
The estimated depth for quantum chemistry is provided in the lower half of Table \ref{tab:scaling_mo_using_Lorentzian_with_linear_QFT}.
The scaling of computational time in this case coincides with that of depth since the encoding is deterministic.

\subsection{Classical computation of finding the optimal LC for a target function}
\label{sed:finding_optimal_LC}

There is often a practical case where we know an $n_q$-qubit state
$
| \psi_{\mathrm{ideal}} \rangle
=
\sum_{j = 0}^{N - 1}
\psi_{\mathrm{ideal}} (x_j) | j \rangle_{n_q}
$
for which we do not, however,
know the optimal parameters (expansion coefficients, decay rates, and centers) of LFs comprising an LC that well approximates the ideal state.
To find the optimal parameters, we introduce a trial state of the form
\begin{align}
    | \psi [\boldsymbol{d}, \boldsymbol{a}, \boldsymbol{k}_{\mathrm{c}} ] \rangle
    =
        \sum_{\ell = 0}^{n_{\mathrm{loc}} - 1}
            d_\ell
            | L; a_\ell, k_{\mathrm{c} \ell} \rangle_{n_q}
    ,
\end{align}
where $\boldsymbol{d}, \boldsymbol{a},$ and
$\boldsymbol{k}_{\mathrm{c}}$ represent the expansion coefficients, decay rates, and centers, respectively, of $n_{\mathrm{loc}}$ LFs.
For a fixed $n_{\mathrm{loc}}$,
we want to find on a classical computer the combination of these parameters so that the squared overlap
$
    F
    (\boldsymbol{d}, \boldsymbol{a}, \boldsymbol{k}_{\mathrm{c}})
    \equiv
        |
        \langle
        \psi
        [\boldsymbol{d}, \boldsymbol{a}, \boldsymbol{k}_{\mathrm{c}}]
        | \psi_{\mathrm{ideal}} \rangle
        |^2
$
between the trial state and the target state is 
maximized with obeying the normalization condition
$\| | \psi [\boldsymbol{d}, \boldsymbol{a}, \boldsymbol{k}_{\mathrm{c}}] \rangle \|^2 = 1.$

There may exist various numerical procedures for finding the optimal parameters.
Here we provide an outline of a possible one that exploits the fact that the objective function $F$ is in a quadratic form with respect to $\boldsymbol{d}.$
Specifically, from
$
g_\ell
\equiv
\sum_{j = 0}^{N - 1}
\psi_{\mathrm{ideal}} (x_j)
L_j (a_\ell, n_q)
$
for each $\ell$ and the overlaps between the LF states (see Appendix \ref{appendix:overlap_btwn_Lorentzian_states}),
we construct the matrices $G$ and $S$ as
$G_{\ell \ell'} \equiv g_\ell g_{\ell'}$ and
$
S_{\ell \ell'}
\equiv 
V
(
a_\ell, a_{\ell'},
k_{\mathrm{c} \ell} - k_{\mathrm{c} \ell'}
),
$
respectively.
The stationarity conditions for $F$ with respect to the expansion coefficients lead to the generalized eigenvalue problem:
$
    G
    \boldsymbol{c}
    =
        \lambda
        S
        \boldsymbol{c}
    ,
$
where $\lambda$ is the Lagrange multiplier responsible for the normalization condition and
$\boldsymbol{c}$ is the solution of this equation.
The solution out of $n_{\mathrm{loc}}$ that maximizes $F$ gives the globally optimal coefficients for fixed $\boldsymbol{a}$ and $\boldsymbol{k}_{\mathrm{c}}:$
\begin{gather}
    \boldsymbol{d}
    (\boldsymbol{a}, \boldsymbol{k}_{\mathrm{c}})
    =
    \argmax_{G \boldsymbol{c} = \lambda S \boldsymbol{c} }
        F
        (\boldsymbol{c}, \boldsymbol{a}, \boldsymbol{k}_{\mathrm{c}})
    .
    \label{gen_loc_state:optimal_coeff_for_lc}
\end{gather}
From these coefficients,
we can define a new objective function
$
F_{\mathrm{d c}}
(\boldsymbol{a}, \boldsymbol{k}_{\mathrm{c}})
\equiv
F
(
    \boldsymbol{d}
    (\boldsymbol{a}, \boldsymbol{k}_{\mathrm{c}})
    ,
    \boldsymbol{a}, \boldsymbol{k}_{\mathrm{c}}
)
.
$
Since the normalization condition for the trial state is already taken into account in this function,
we should find the optimal $\boldsymbol{a}$ and $\boldsymbol{k}_{\mathrm{c}}$ by maximizing
$F_{\mathrm{d c}}$ with no constraint.

For fixed $\boldsymbol{k}_{\mathrm{c}},$
we can obtain the optimal decay rates
$\boldsymbol{a} (\boldsymbol{k}_{\mathrm{c}})$
that maximize $F_{\mathrm{d c}}$ via some algorithm for continuous variables.
We can then define yet another objective function
$
F_{\mathrm{c}} (\boldsymbol{k}_{\mathrm{c}})
\equiv
F_{\mathrm{dc}}
(
\boldsymbol{a} (\boldsymbol{k}_{\mathrm{c}}),
\boldsymbol{k}_{\mathrm{c}}
)
,
$
a function only of the discrete variables.
The optimal centers $\boldsymbol{k}_{\mathrm{c}}$ can be obtained, for example, via the Metropolis method.
More specifically,
whether the change in the center of each LF (moved by $+1$ or moved by $-1$ or not moved) at each step is decided based on a random number and
the value of $F_{\mathrm{c}}$ that would be taken if it is moved.
The pseudocodes for such a procedure are provided in
Appendix \ref{appendix:pseudocodes_for_optimal_LC}.

Let us consider the classical-computational time for finding the optimal coefficients given by Eq.~(\ref{gen_loc_state:optimal_coeff_for_lc}).
Its scaling in terms of $n_{\mathrm{loc}}$ and the number $N$ of grid points can be estimated as follows.
Since the computational time for evaluating $g_\ell$ for each $\ell$ is clearly $\mathcal{O} (N)$ from its definition,
that for constructing $G$ is
$\mathcal{O} (\max (n_{\mathrm{loc}}^2, n_{\mathrm{loc}} N)).$
The computational time for constructing $S$ is $\mathcal{O} (n_{\mathrm{loc}}^2).$
After the construction of $G$ and $S,$ the generalized eigenvalue problem is solved via a standard method by spending time of $\mathcal{O} (n_{\mathrm{loc}}^3).$
The computational time for finding the optimal coefficients is thus
$
\mathrm{time}
( \boldsymbol{d} (\boldsymbol{a}, \boldsymbol{k}_{\mathrm{c}}) )
=
\mathcal{O} (\max (n_{\mathrm{loc}}^3, n_{\mathrm{loc}} N)).
$
If the cubic scaling in $n_{\mathrm{loc}}$ for full diagonalization is problematic in a practical case,
one can resort to a more rapid optimizer such as the conjugate gradient method that finds a local optimum.
We do not mention here the computational time spent by the optimization methods for $\boldsymbol{a}$ and $\boldsymbol{k}_{\mathrm{c}}$
since that differs according to the methods.

\begin{figure}
\begin{center}
\includegraphics[width=7cm]{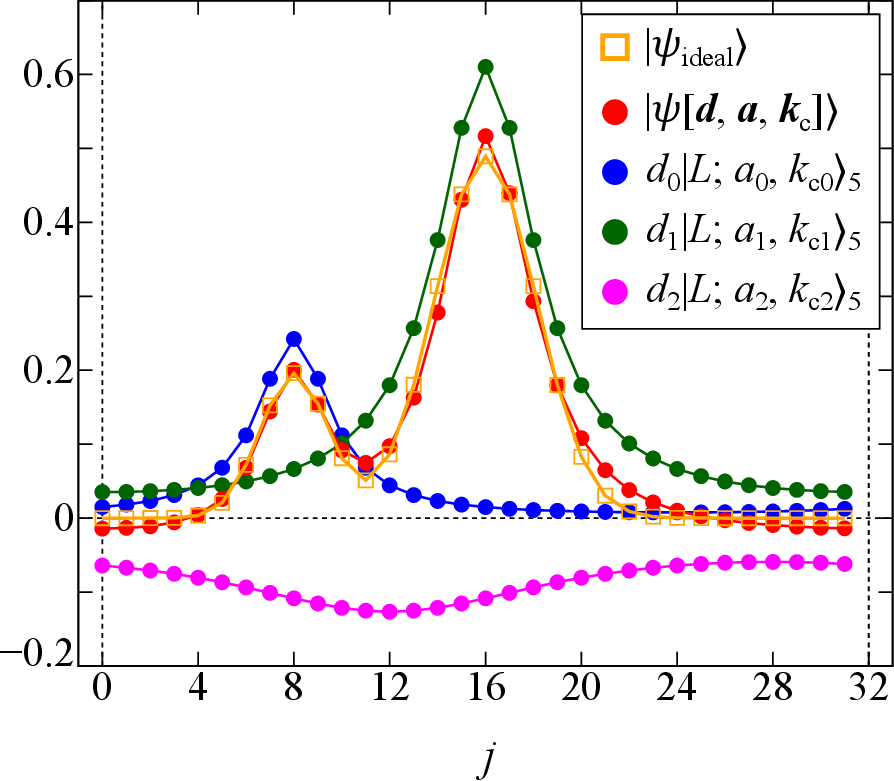}
\end{center}
\caption{
Target state $| \psi_{\mathrm{ideal}} \rangle$ for $n_q = 5$ data qubits and the LC
$| \psi [\boldsymbol{d}, \boldsymbol{a}, \boldsymbol{k}_{\mathrm{c}}] \rangle$
of LF states ($n_{\mathrm{loc}} = 3$)
that optimally approximates the target state.
The contribution from each LF state is also shown.
}
\label{fig:lc_example}
\end{figure}

As an example for the procedure described above,
we tried to find the optimal LC to express the sum of two Gaussian functions:
$
    \psi_{\mathrm{ideal}} (x_j)
    \propto
        \exp
        ( -(j - 16)^2/9 )
        +
        0.4
        \exp
        ( -(j - 8)^2/4 )
$
at $j = 0, \dots, 31$ for $n_q = 5$ data qubits by using $n_{\mathrm{loc}} = 3$ LFs.
We obtained the decay rates
$a_0 = 0.360, a_1 = 0.490,$ and $a_2 = 1.672$
and the centers
$k_{\mathrm{c} 0} = 8, k_{\mathrm{c} 1} = 16,$ and $k_{\mathrm{c} 2} = 12$ for the LFs.
Their coefficients were 
$d_0 = 0.417, d_1 = 1.23,$ and $d_2 = -0.507.$
The target state $| \psi_{\mathrm{ideal}} \rangle$ and
the optimal state $| \psi [\boldsymbol{d}, \boldsymbol{a}, \boldsymbol{k}_{\mathrm{c}}] \rangle$
are plotted in Fig.~\ref{fig:lc_example}.
The squared overlap between the target and optimal states was found to be $0.992.$

\section{Demonstrations using real quantum computers}

To confirm the validity of our encoding techniques,
we generated LCs of LFs on the real quantum computers provided by IBM \cite{ibmq_quantum}.
We used qiskit \cite{qiskit2024} for the compilation and execution of the circuits for the cloud quantum computing platform.
We adopted the dynamical decoupling technique to suppress errors during the execution of quantum circuits {\cite{bib:6102}}.
We did not employ QARA since we wanted to perform quantum computation using as few real qubits as possible.

\begin{figure*}
\begin{center}
\includegraphics[width=16cm]{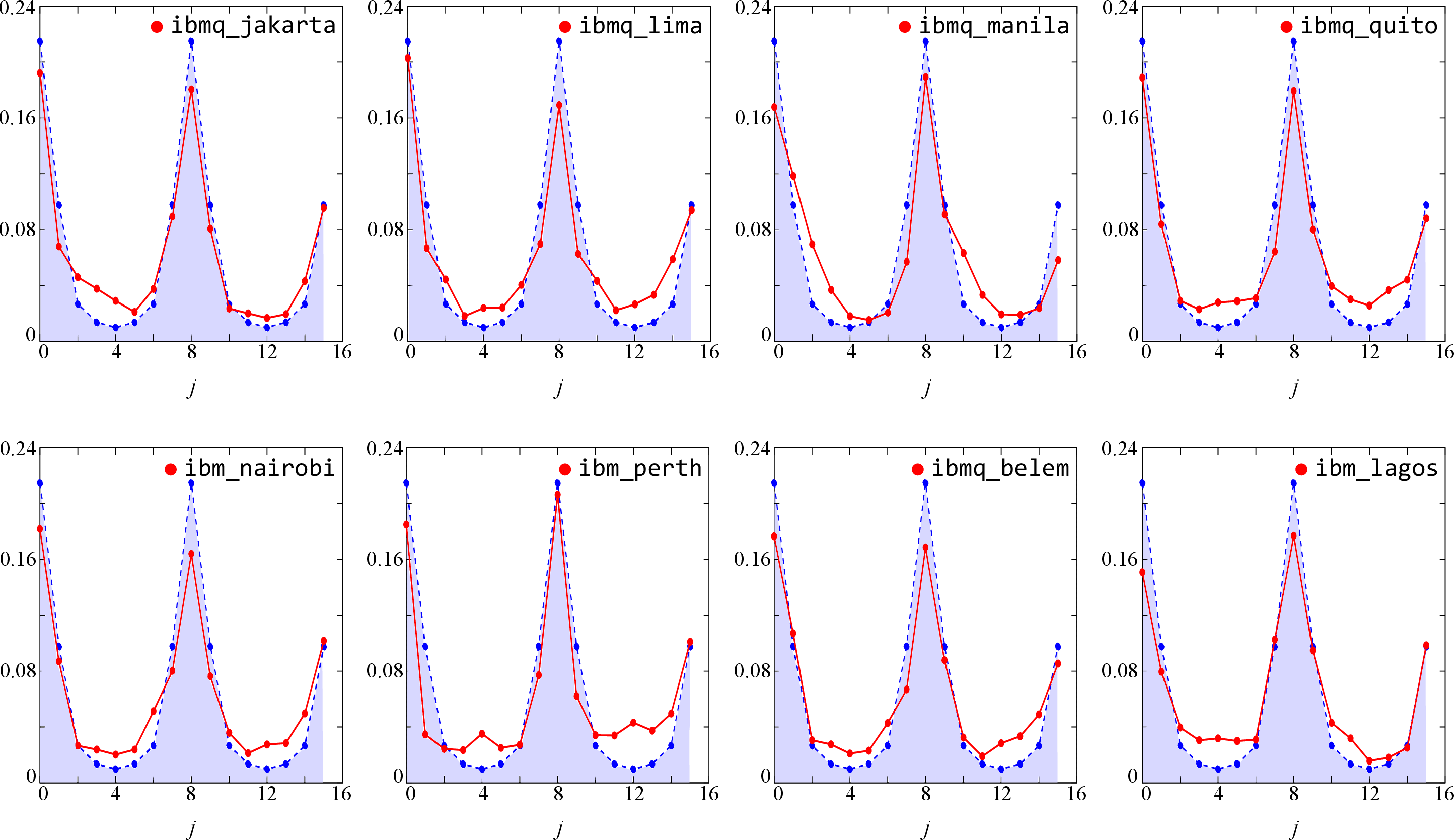}
\end{center}
\caption{
Observed probability distributions
$| \langle j |_{n_q} |\psi_{\mathrm{lc}} \rangle |^2$
as red circles according to probabilistic encoding of 
$
| \psi_{\mathrm{lc}} \rangle
\propto
| L; 0.5, 0 \rangle_{n_q} + | L; 0.5, 8 \rangle_{n_q}
$
for $n_q = 4$ qubits on the real quantum computers.
The ideal distribution, which is of course common to all the machines,
is also shown as blue circles on each panel.
}
\label{fig:ibm_different_machines}
\end{figure*}

Figure \ref{fig:ibm_different_machines} shows the observed probability distributions for $n_q = 4$ data qubits and one ancilla according to the probabilistic encoding of 
$
| \psi_{\mathrm{lc}} \rangle
\propto
| L; a_0, k_{\mathrm{c} 0} \rangle_{n_q}
+
| L; a_1, k_{\mathrm{c} 1} \rangle_{n_q}
$
with $a_0 = a_1 = 0.5, k_{\mathrm{c} 0} = 0,$
and $k_{\mathrm{c} 1} = 8.$
The ideal distribution is also shown on each panel.
It is seen that the observed distributions and the ideal one are in reasonable agreement for all the eight quantum computers.
The number of CNOT gates in the transpiled circuit was 23.

\begin{figure*}
\begin{center}
\includegraphics[width=14cm]{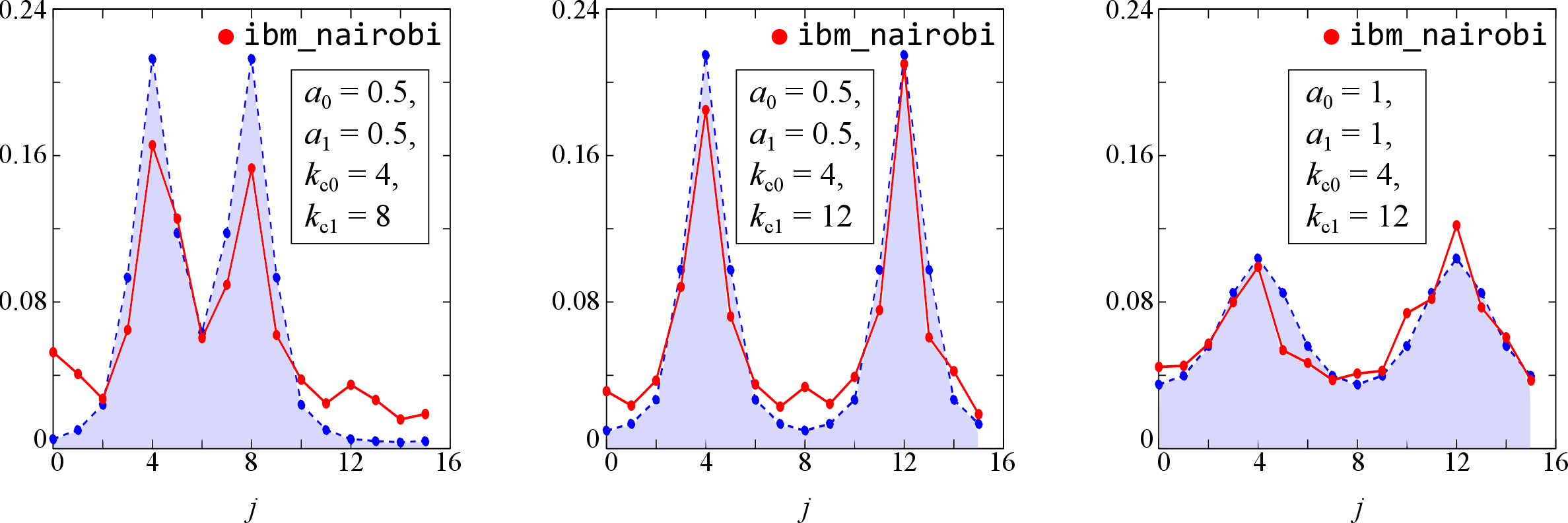}
\end{center}
\caption{
Observed probability distributions as red circles according to probabilistic encoding of 
$
| \psi_{\mathrm{lc}} \rangle
\propto
| L; a_0, k_{\mathrm{c} 0} \rangle_{n_q}
+
| L; a_1, k_{\mathrm{c} 1} \rangle_{n_q}
$
with various combinations of 
decay rates and centers for $n_q = 4$ qubits on ibm\_nairobi.
The ideal distributions are also shown as blue circles.
}
\label{fig:ibm_nairobi_various_params}
\end{figure*}

To see how the differences between the shapes of ideal functions affect the quality of resultant states,
we tried three combinations of decay rates and centers of two LFs for ibm\_nairobi.
The results are shown in
Fig.~\ref{fig:ibm_nairobi_various_params}.
The numbers of CNOT gates in the transpiled circuits for the panels from the left were 31, 23, and 29.

\section{Conclusions}

In summary, we developed a probabilistic encoding technique that generates an LC of LFs with arbitrary complex coefficients.
Since the ancillae increase only logarithmically with respect to the number of constituent LFs regardless of the size of the data register and
QFT is performed only once,
this technique achieves efficient encoding.
Furthermore, the QARA technique was shown to determinize the encoding technique with controllable errors.
We performed the probabilistic encoding technique on commercially available real quantum computers and found that the results were in reasonable agreement with the theoretical ones.

Looking beyond the noisy intermediate-scale quantum device (NISQ) era,
one of the intriguing applications of our generic scheme will be the encoding of MOs used in quantum chemistry in real space.
In particular, we found via resource estimation that the computational time for encoding a localized MO is polynomial in terms of the logarithm of the electron number.
These insights are encouraging since encoding MOs is crucial for the state preparation of an initial many-electron wave function.
The encoding techniques developed in the present study thus make quantum chemistry in real space more promising on fault-tolerant quantum computers.
Also, the techniques will be a powerful tool in the field of financial analyses using quantum computers,
where they often want to encode various probability distribution functions in continuous space efficiently.

\begin{acknowledgments}
We thank Xinchi Huang for fruitful discussion. 
This work was supported by JSPS KAKENHI under Grant-in-Aid for Scientific Research
No. 21H04553,
No. 20H00340,
and No. 22H01517,
JSPS KAKENHI under Grant-in-Aid for Transformative Research Areas
No. JP22H05114,
JSPS KAKENHI under Grant-in-Aid for EarlyCareer Scientists
No. JP24K16985.
This study was partially carried out using the TSUBAME3.0 supercomputer at the Tokyo Institute of Technology and the facilities of the Supercomputer Center,
the Institute for Solid State Physics,
the University of Tokyo.
The author acknowledges the contributions and discussions provided by the members of Quemix Inc.
This work was partially supported by the Center of Innovations for Sustainable Quantum AI (JST Grant Number JPMJPF2221).
We acknowledge the use of IBM Quantum services for this work.
\end{acknowledgments}

\begin{widetext}

\appendix

\section{Error analysis for QARA}
\label{sec:error_analysis_for_ampl_red}

Let us consider a situation where the observer has an inaccurate estimation of the weight of the success state:
$w^{(\mathrm{est})} = w + \Delta w,$
where $w$ is the true value and $\Delta w$ is an error having intruded into the QAE process.
The observer then determines an iteration number
\begin{align}
    m_{\mathrm{opt}}'
    =
        \left\lceil
        \frac{\pi}{4 \theta_{w^{(\mathrm{est})}}}
        -
        \frac{1}{2}
        \right\rceil
    \label{gen_loc_state:m_opt_for_QAA_with_QAE_error}
\end{align}
instead of Eq.~(\ref{gen_loc_state:m_opt_for_QAA}),
without being aware of the error.
Furthermore,
the observer adopts an inaccurate circuit parameter
\begin{align}
    \theta_{\mathrm{AR,opt}}'
    =
        \arccos
        \left(
            \frac{1}{\sqrt{w^{(\mathrm{est})}}}
            \sin \frac{\pi}{4 m_{\mathrm{opt}}' + 2}
        \right)
    \label{gen_loc_state:angle_for_determinization_with_QAE_error}
\end{align}
for the amplitude reduction instead of
Eq.~(\ref{gen_loc_state:angle_for_determinization}).
Since we are interested in amplitude amplification for a very small weight of the success state,
we assume $| \Delta w | < w \ll 1$ in what follows.
The angle corresponding to the erroneous weight is expressed as
\begin{align}
    \theta_{w^{(\mathrm{est})}}
    &=
        \arcsin \sqrt{w}
        +
        \frac{1}{2}
        \sqrt{\frac{w}{1 - w}}
        \frac{\Delta w}{w}
        -
        \frac{(1 - 2 w) \sqrt{w}}{8 (1 - w)^{3/2}}
        \left(
            \frac{\Delta w}{w}
        \right)^2
        +
            \mathcal{O}
            \left(
            \left( \frac{\Delta w}{w} \right)^3
            \right)
    ,
    \label{ampl_red_and_ampl:theta_w_plus_delta_w}
\end{align}
where we used the Taylor expansion of arcsin function.
By expanding 
Eq.~(\ref{gen_loc_state:m_opt_for_QAA_with_QAE_error})
in $\Delta w/w,$ we obtain
\begin{align}
    m_{\mathrm{opt}}'
    &\approx
        \frac{\pi}{4 \theta_{w^{(\mathrm{est})}}}
    \nonumber \\
    &=
        \frac{\pi}{4 \arcsin \sqrt{w}}
        \left(
            1
            +
            \mu_1 (w)
            \frac{\Delta w}{w}
            +
            \mu_2 (w)
            \left(
                \frac{\Delta w}{w}
            \right)^2
        \right)
        +
            \mathcal{O}
            \left(
            \left( \frac{\Delta w}{w} \right)^3
            \right)
    ,
\end{align}
where we defined
\begin{align}
    \mu_1 (w)
    &\equiv
        -
        \frac{\sqrt{w}}{2 \sqrt{1 - w} \arcsin \sqrt{w}}
\end{align}
and
\begin{align}
    \mu_2 (w)
    &\equiv
        \frac{(1 - 2 w) \sqrt{w}}{8 (1 - w)^{3/2} \arcsin \sqrt{w}}
        +
        \frac{w}{4 (1 - w) (\arcsin \sqrt{w})^2}
    .
\end{align}
The weight $w_{\mathrm{AR}}'$ of success state just after the amplitude reduction in this case is calculated 
by substituting the circuit parameter in
Eq.~(\ref{gen_loc_state:angle_for_determinization_with_QAE_error})
into
Eq.~(\ref{gen_loc_state:success_prob_after_ampl_reduction}).
Its square root is calculated, via tedious manipulation of expressions, as
\begin{align}
    \sqrt{w_{\mathrm{AR}}'}
    =
    \sqrt{
        w
        \cos^2 \theta_{\mathrm{AR,opt}}'
    }
    \approx
        \sqrt{\frac{w}{ w^{(\mathrm{est})} }}
        \sin
        \frac{\pi}{4 m_{\mathrm{opt}}'}
    =
        \sqrt{w}
        +
        \mathcal{O}
        \left(
        \left( \frac{\Delta w}{w} \right)^3
        \right)
    .
\end{align}
The angle corresponding to $w_{\mathrm{AR}}'$ is thus
\begin{align}
    \theta_{ w_{\mathrm{AR}}' }
    =
        \arcsin
        \sqrt{w_{\mathrm{AR}}' }
    \approx
        \arcsin
        \sqrt{w}
        +
        \mathcal{O}
        \left(
        \left( \frac{\Delta w}{w} \right)^3
        \right)
    .
\end{align}
The weight of success state after the completion of the amplification process is thus calculated as
\begin{align}
    W_{\mathrm{ARA}}
    &=
        \sin^2
        \left(
            (2 m_{\mathrm{opt}}' + 1)
            \theta_{ w_{\mathrm{AR}}' }
        \right)
    \nonumber \\
    &\approx
        \sin^2
        \left(
            2 m_{\mathrm{opt}}'
            \theta_{ w_{\mathrm{AR}}' }
        \right)
    \nonumber \\
    &\approx
        \sin^2
        \left(
        \frac{\pi}{2}
        +
        \frac{\pi}{2} \mu_1 (w)
        \frac{\Delta w}{w}
        +
        \frac{\pi}{2} \mu_2 (w)
        \left( \frac{\Delta w}{w} \right)^2
        \right)
    .
\end{align}
In order for the weight of failure state,
$1 - W_{\mathrm{ARA}},$ to be smaller than $\epsilon,$
the condition for $\Delta w/w$ up to the first order reads
\begin{align}
        \frac{\pi}{2} \mu_1 (w)
        \frac{| \Delta w |}{w}
        <
        \arcsin \sqrt{\epsilon}
    .
\end{align}
When we approach the limit of $w \to 0$ with keeping $\Delta w/w$ finite,
the condition above becomes
\begin{align}
        \frac{| \Delta w |}{w}
        <
        \frac{4}{\pi}
        \arcsin \sqrt{\epsilon}
\end{align}
since $\mu_1 (w \to 0) = -1/2.$
The equation above also gives an approximate error as a function of $\Delta w/w:$
\begin{align}
    \epsilon_{\mathrm{QARA}}
    =
        \sin^2 \frac{\pi \Delta w}{4 w}
        .
\end{align}

\section{Implementation of $U_{\mathrm{shift}} (k)$}
\label{appendix:impl_of_phase_shift}

For a computational basis
$
| j \rangle_{n_q} = 
| j_{n_q - 1} \rangle \otimes \cdots \otimes | j_0 \rangle
\
(j_0, \dots, j_{n_q - 1} = 0, 1)
$
with $j = \sum_{\ell = 0}^{n_q - 1} 2^\ell j_{\ell},$
the action of $U_{\mathrm{shift}} (k)$ on it is written, from
Eq.~(\ref{gen_loc_state:def_U_shift}),
as
\begin{align}
    U_{\mathrm{shift}} (k)
    | j \rangle_{n_q}
    &=
        \exp
        \left(
            -i
            \frac{2 \pi k}{N}
            j
        \right)
        | j \rangle_{n_q}
    \nonumber \\
    &=
        \prod_{m = 0}^{n_q - 1}
        \exp
        \left(
            -i
            \frac{2 \pi k}{N}
                2^m
                j_m
        \right)
        | j \rangle_{n_q}
    \nonumber \\
    &=
        Z (\phi_{n_q - 1})
        | j_{n_q - 1} \rangle
        \otimes
        \cdots
        \otimes
        Z (\phi_0)
        | j_0 \rangle
    .
\end{align}
$\phi_m \equiv -2 \pi k 2^m/N$ is the angle parameter for the single-qubit phase shift
$Z (\phi) \equiv \mathrm{diag} (1, e^{i \phi}).$
The above equation indicates that
$U_{\mathrm{shift}} (k)$ can be implemented as $n_q$ single-qubit unitaries performed simultaneously.

\section{Efficient implementation of multiply controlled multiple single-qubit unitaries}
\label{sec:impl_for_multi_ctrl_multi_1q}

Here we describe techniques for efficient implementation of multiply controlled multiple single-qubit (MCM1) unitaries
by generalizing the phase gradient method \cite{phase_gradient_circuits}.

\subsection{$\mathrm{C}^{m_{\mathrm{c}}} X^{\otimes m_{\mathrm{t}}}$}
\label{sec:impl_for_multi_ctrl_X_gates}

We first consider $X$ gates on $m_{\mathrm{t}}$ target qubits $(m_{\mathrm{t}} \geq 2)$ controlled by $m_{\mathrm{c}}$ qubits,
as depicted on the left-hand side of 
Fig.~\ref{fig:circuit_multi_ctrl_multi_1q}(a).
This circuit is known as the quantum fan-out gate \cite{bib:6606, bib:6607}.
We adopt the same technique for this $(m_{\mathrm{c}} + m_{\mathrm{t}})$-qubit circuit as in the original method \cite{phase_gradient_circuits}.
We describe the technique here in order for this paper to be self contained.

We define partial circuits
$S_k$ for $k = 1, \dots, \lceil \log_2 m_{\mathrm{t}} \rceil$ as shown on the right-hand side of
Fig.~\ref{fig:circuit_multi_ctrl_multi_1q}(a).
$S_k$ for each $k$ is composed only of 
at most $2^{k - 1}$ simultaneously executable CNOT gates.
One can easily confirm
the equality in Fig.~\ref{fig:circuit_multi_ctrl_multi_1q}(a),
that is, $\mathrm{C}^{m_{\mathrm{c}}} X^{\otimes m_{\mathrm{t}}}$ can be implemented by
arranging the $S_k$ circuits in descending order,
appending a $\mathrm{C}^{m_{\mathrm{c}}} X$ gate,
and arranging the $S_k$ circuits in ascending order.
Since the $\mathrm{C}^{m_{\mathrm{c}}} X$ operation admits a linear-depth implementation \cite{bib:5614, bib:5662, bib:5995},
the total depth is given by
\begin{align}
    \mathrm{depth}
    (
        \mathrm{C}^{m_{\mathrm{c}}}
        X^{\otimes m_{\mathrm{t}}}
    )
    &=
        \mathcal{O}
        \left(
            \max
            \left(
            m_{\mathrm{c}},
            \log m_{\mathrm{t}}
            \right)
        \right)
    .
\end{align}

\subsection{MCM1 unitaries}

Let us consider a case where the $n_q$ qubits in the data register undergo generic single-qubit unitaries $U_\ell \ (\ell = 0, \dots, n_q - 1)$
simultaneously controlled by the $n_{\mathrm{A}}$ ancillae,
as depicted on the left-hand side of
Fig.~\ref{fig:circuit_multi_ctrl_multi_1q}(b).
We demonstrate below that this operation can be implemented with a logarithmic depth in terms of $n_q.$

$U_\ell$ for each $\ell$ can be expressed as a rotation 
$
e^{i \alpha_\ell}
e^{-i \boldsymbol{n}_\ell \cdot \boldsymbol{\sigma} \theta_\ell/2}
$
by using a phase factor $e^{i \alpha_\ell},$
a rotation axis $\boldsymbol{n}_\ell,$
and a rotation angle $\theta_\ell$ \cite{Nielsen_and_Chuang}.
By using the polar angle $\vartheta_\ell$ and the azimuthal angle $\varphi_\ell$ of the rotation axis,
we define a single-qubit operator
\begin{align}
    V_\ell
    \equiv
        \begin{pmatrix}
            \cos (\vartheta_\ell/2) & \sin (\vartheta_\ell/2) \\
            e^{i \varphi_\ell} \sin (\vartheta_\ell/2)
            &
            -e^{i \varphi_\ell} \cos (\vartheta_\ell/2)
        \end{pmatrix}
    \label{sec:impl_of_MCM1_def_V}
\end{align}
and a diagonal operator
$D_\ell \equiv e^{i (\alpha_\ell - \theta_\ell/2)} Z (\theta_\ell),$
where 
$Z (\theta) \equiv \mathrm{diag} (1, e^{i \theta})$
is the phase shift unitary.
With them, we diagonalize the unitary for the $\ell$th data qubit as
$U_\ell = V_\ell D_\ell V_\ell^\dagger$
and construct the circuit $\mathcal{C}$ depicted on the right-hand side of
Fig.~\ref{fig:circuit_multi_ctrl_multi_1q}(b),
where we defined $\widetilde{\alpha} \equiv \sum_{\ell = 0}^{n_q - 1} \alpha_\ell$ and
$Z_\ell \equiv Z (\theta_\ell).$
It is obvious that the left and right circuits are equivalent to each other when the state of any one of the ancillae is $| 0 \rangle.$

Let us check whether $\mathcal{C}$ works the same way as the left circuit when the ancillary state is $| 1 \rangle^{\otimes n_{\mathrm{A}}}.$
The state of the entire system involving an arbitrary state 
$| j \rangle_{n_q} \equiv | j_{n_q - 1} \rangle \otimes \cdots \otimes | j_0 \rangle$
of the data register undergoes the circuit as
\begin{align}
    | j \rangle_{n_q} 
    \otimes
    | 1 \rangle^{\otimes n_{\mathrm{A}}}
    &\xmapsto{Z (\widetilde{\alpha}), V^\dagger, \sigma_x, Z^{1/2 \dagger}, \sigma_x,  Z^{1/2}, V}
        \left(
            \bigotimes_{\ell = 0}^{n_q - 1}
                V_\ell
                Z_\ell^{1/2}
                \sigma_x
                Z_\ell^{1/2 \dagger}
                \sigma_x
                V_\ell^\dagger
                | j_\ell \rangle
        \right)
        \otimes
        e^{i \widetilde{\alpha}}
        | 1 \rangle^{\otimes n_{\mathrm{A}}}
    \nonumber \\
    &=
        \left(
            \bigotimes_{\ell = 0}^{n_q - 1}
                e^{-i \theta_\ell/2}
                e^{-i (\alpha_\ell - \theta_\ell/2)}
                U_\ell
                | j_\ell \rangle
        \right)
        \otimes
        e^{i \widetilde{\alpha}}
        | 1 \rangle^{\otimes n_{\mathrm{A}}}
    \nonumber \\
    &=
        \left(
            \bigotimes_{\ell = 0}^{n_q - 1}
                U_\ell
                | j_\ell \rangle
        \right)
        \otimes
        | 1 \rangle^{\otimes n_{\mathrm{A}}}
    ,
    \label{sec:impl_of_MCM1_action_of_circuit}
\end{align}
where we used the relation
$
\sigma_x Z_\ell^{1/2 \dagger} \sigma_x
=
e^{-i \theta_\ell/2} Z_\ell^{1/2}
.
$
Equation (\ref{sec:impl_of_MCM1_action_of_circuit}) confirms that the equality in
Fig.~\ref{fig:circuit_multi_ctrl_multi_1q}(b)
holds regardless of the ancillary state.
The only many-qubit operations in $\mathcal{C}$ are
$\mathrm{C}^{n_{\mathrm{A}} - 1} Z (\widetilde{\alpha})$
and
$\mathrm{C}^{n_{\mathrm{A}}} X^{\otimes n_q}.$
The former can be implemented with a linear depth \cite{bib:5614, bib:5662, bib:5995} and
the latter can be implemented using the technique shown in 
Fig.~\ref{fig:circuit_multi_ctrl_multi_1q}(a).
The total depth for the MCM1 circuit is thus
\begin{align}
    \mathrm{depth} (\mathcal{C})
    =
        \mathcal{O}
        \left(
            \max
            ( n_{\mathrm{A}}, \log n_q )
        \right)
        .
\end{align}

For a special case where each of the rotation axes is along the $z$ axis and $\varphi_\ell$ is set to $\pi,$
$V_\ell$ in Eq.~(\ref{sec:impl_of_MCM1_def_V}) is identity.
$\mathcal{C}$ in such a case is the circuit in the original phase gradient method \cite{phase_gradient_circuits}.

\begin{figure}
\begin{center}
\includegraphics[width=14cm]{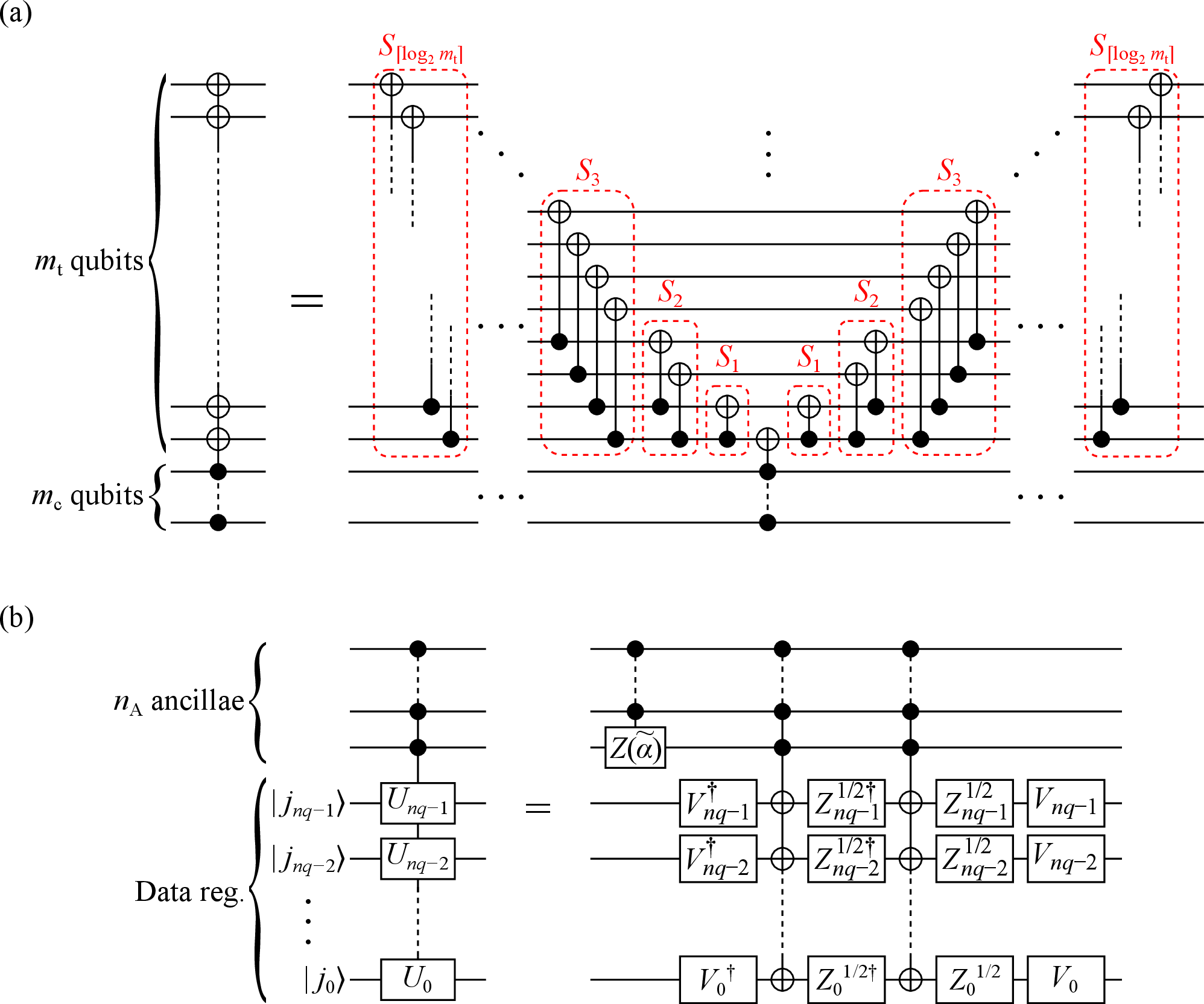}
\end{center}
\caption{
(a)
$\mathrm{C}^{m_{\mathrm{c}}} X^{\otimes m_{\mathrm{t}}}$ operation can be implemented by arranging the $S_k$ circuits and a $\mathrm{C}^{m_{\mathrm{c}}} X$ gate.
(b)
MCM1 unitaries consisting of generic single-qubit unitaries can be implemented with a logarithmic depth in terms of $n_q.$
}
\label{fig:circuit_multi_ctrl_multi_1q}
\end{figure}

\section{Formulation for an LC of complex localized functions via complex coefficients}
\label{appendix:complex_funcs_and_complex_coeffs}

Let us consider a case where we want to encode an LC of localized complex functions
$\{ f_\ell \}_{\ell = 0}^{n_{\mathrm{loc}} - 1}$
via complex coefficients
$\{ d_\ell \}_{\ell = 0}^{n_{\mathrm{loc}} - 1}.$
By separating the real and imaginary parts,
we write 
$f_\ell = f^{(\mathrm{R})}_\ell + i f^{(\mathrm{I})}_\ell$
and
$d_\ell = | d_\ell | e^{i \vartheta_\ell}$
for each $\ell.$
We assume that the unitaries
$U^{(\ell, \mathrm{R})}$ and $U^{(\ell, \mathrm{I})}$
that generate
$f^{(\mathrm{R})}_\ell$ and $f^{(\mathrm{I})}_\ell,$
respectively, at the origin are known.

We define 
$
\widetilde{U}^{(\ell, \mathrm{R})}
\equiv
U_{\mathrm{shift}}^{(\ell)}
\cdot
\mathcal{F}_{\mathrm{quant}}^\dagger
\cdot
e^{i \vartheta_\ell}
U_{\mathrm{orig}}^{(\ell, \mathrm{R})}
$
and
$
\widetilde{U}^{(\ell, \mathrm{I})}
\equiv
U_{\mathrm{shift}}^{(\ell)}
\cdot
\mathcal{F}_{\mathrm{quant}}^\dagger
\cdot
i e^{i \vartheta_\ell}
U_{\mathrm{orig}}^{(\ell, \mathrm{I})}
.
$
for each $\ell.$
By using the states $\{ | \widetilde{f}_\ell \rangle \}_\ell$
representing the displaced complex functions,
the desired state is written as 
\begin{align}
    | \psi_{\mathrm{lc}} \rangle
    &=
        \sum_{\ell = 0}^{n_{\mathrm{loc}} - 1}
            d_\ell
            | \widetilde{f}_\ell \rangle
    \nonumber \\
    &=
        \sum_{\ell = 0}^{n_{\mathrm{loc}} - 1}
            | d_\ell | e^{i \vartheta_\ell}
            T (k_{\mathrm{c} \ell})
            \left(
                U_{\mathrm{orig}}^{(\ell, \mathrm{R})}
                +
                i
                U_{\mathrm{orig}}^{(\ell, \mathrm{I})}
            \right)
            | 0 \rangle_{n_q}    
    \nonumber \\
    &=
        \mathcal{F}_{\mathrm{quant}}
        \sum_{\ell = 0}^{n_{\mathrm{loc}} - 1}
            \left(
                | d_\ell |
                \widetilde{U}^{(\ell, \mathrm{R})}
                +
                | d_\ell |
                \widetilde{U}^{(\ell, \mathrm{I})}
            \right)
            | 0 \rangle_{n_q}
    ,
\end{align}
where $T$ is the translation operator defined in 
Eq.~(\ref{gen_loc_state:def_translation_opr}).
The summation on the right-hand side of equation above is an LC of $2 n_{\mathrm{loc}}$ unitaries with real coefficients.
This means that probabilistic encoding of 
$| \psi_{\mathrm{lc}} \rangle$ is possible by adopting
the protocol for real values in the main text with small modifications for $n_{\mathrm{A}} + 1$ ancillae.

\section{Extension to multidimensional cases}
\label{appendix:multidimensional_space}

Here we establish a straightforward extension of the encoding scheme for one dimension to multidimensional cases.
Although the extension is possible for arbitrary dimension,
we focus on three-dimensional (3D) cases for practical interest.

\subsection{For generic basis functions}

We introduce $x, y,$ and $z$ registers each of which is built up of $n_q$ qubits,
forming the $3 n_q$-qubit data register. 
We discretize each of the three directions for a cube into $N \equiv 2^{n_q}$ grid points similarly to the one-dimensional case.
We assume the spacing $\Delta x$ of the grid points in each direction to be common.
We want to encode a function in the 3D space by using displaced localized function.
To this end,
by adopting $n_f$ localized functions for each direction,
we build a basis set from product basis functions
\begin{align}
    f_{\ell} (x, y, z)
    \equiv
        f_{b_{\ell x}} (x; a_{\ell x})
        f_{b_{\ell y}} (y; a_{\ell y})
        f_{b_{\ell z}} (z; a_{\ell z})
    ,
\end{align}
where
$b_{\ell \mu} = 0, \dots, n_f - 1 \ (\mu = x, y, z)$
specifies the localized function in $\mu$ space for the $\ell$th basis function in the 3D space.
The number of basis functions is thus
$n_{\mathrm{loc}} = n_f^3.$
$a_{\ell \mu}$ represents the parameters in the localized function for the $\mu$ space collectively.
We express the desired state as an LC of these functions as
\begin{align}
    | \psi_{\mathrm{lc}} \rangle
    =
        \sum_{\boldsymbol{j}}
        \sum_{\ell = 0}^{n_{\mathrm{loc}} - 1}
            d_{\ell}
            f_{\ell}
            (
            x_{j_x}
            -
            k_{\mathrm{c} \ell x} 
            \Delta x
            ,
            y_{j_y}
            -
            k_{\mathrm{c} \ell y} 
            \Delta x
            ,
            z_{j_z}
            -
            k_{\mathrm{c} \ell z} 
            \Delta x
            )
            | j_x \rangle_{n_q}
            \otimes
            | j_y \rangle_{n_q}
            \otimes
            | j_z \rangle_{n_q}
    ,
\end{align}
where $k_{\mathrm{c} \ell \mu}$ is the integer coordinate of the center of the $\ell$th basis function.
$x_j = y_j = z_j = j \Delta x$
is the coordinate of the $j$th grid point in each direction.

Since each of the basis functions is separable,
the unitary for generating the single basis function at the origin is obviously
$
U_{\mathrm{orig}}^{(\ell)}
=
U_{\mathrm{orig}}^{(b_{\ell x})}
U_{\mathrm{orig}}^{(b_{\ell y})}
U_{\mathrm{orig}}^{(b_{\ell z})}
,
$
where
$U_{\mathrm{orig}}^{(b_{\ell \mu})} \ (\mu = x, y, z)$
is the $n_q$-qubit unitary
that acts only on the $\mu$ register to generate $f_{b_{\ell \mu}}$ at the origin.
Also, the separability allows us to implement the phase shift operators 
$U_{\mathrm{shift}}^{(\ell, \mu)}$ for the center coordinates
as acting on the $x, y,$ and $z$ registers separately.

\subsection{Probabilistic encoding using LFs as basis functions}

The circuit for probabilistic encoding of $| \psi_{\mathrm{lc}} \rangle$ by using the product bases of discrete LFs is now constructed straightforwardly,
as shown in Fig.~\ref{fig:circuit_lc_3d_Lorentzian_prob}.
This circuit is a 3D extension of $\mathcal{C}_{\mathrm{lc}}^{(\mathrm{L})}$ in
Fig.~\ref{fig:circuit_lc_Lorentzian_prob}(b).
Although the nominal number of the constituent qubits in the data register is $3 n_q$ (instead of $n_q$ in the main text) in this case,
the discussion on the scaling of computational cost in the main text will suffer from no change.

\begin{figure*}
\begin{center}
\includegraphics[width=17cm]{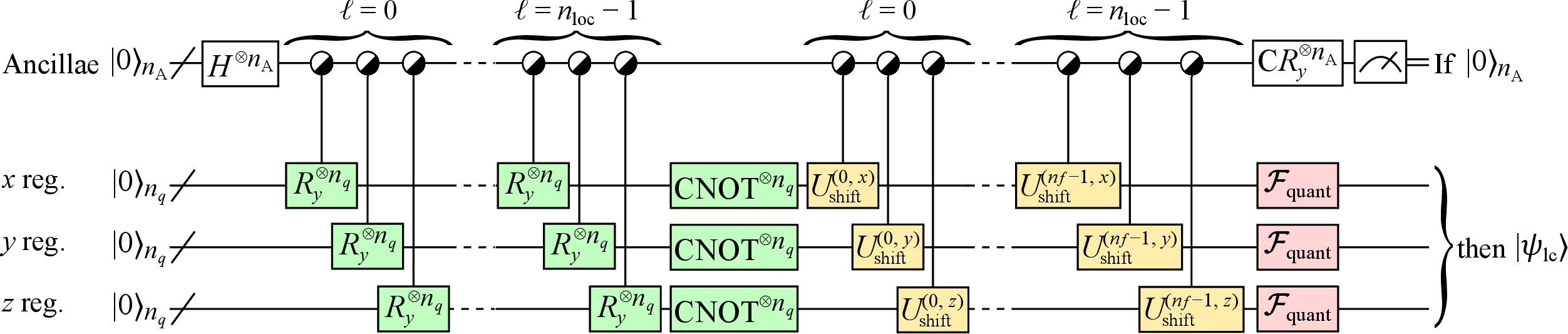}
\end{center}
\caption{
Circuit for probabilistic encoding of the LC of three-LF product basis functions.
}
\label{fig:circuit_lc_3d_Lorentzian_prob}
\end{figure*}

\section{Action of $U^{(\mathrm{S})}$}
\label{appendix:action_of_U_S}

The $n_q$-qubit state
$| q_{n_q - 1} \rangle \otimes \cdots \otimes | q_0 \rangle$
with $q_0, \dots, q_{n_q - 1} = 0, 1$ is the computational basis
$| \sum_{\ell = 0}^{n_q - 1} 2^\ell q_\ell \rangle_{n_q}.$
As confirmed from Fig.~\ref{fig:Slater_and_Lorentzian}(a), 
the qubits other than the $(n_q - 1)$th one undergo the $y$ rotations in $U^{(\mathrm{S})}$ as
\begin{align}
    \overbrace{
        | 0 \rangle
    }^{| q_{n_q - 2} \rangle}
    \otimes
    \cdots
    \otimes
    \overbrace{
        | 0 \rangle
    }^{| q_0 \rangle}
    &\xmapsto{R_y}
        \underbrace{
            \cos \theta_{n_q - 2} \cdots \cos \theta_0
        }_{\equiv C}
        \left(
            | 0 \rangle
            +
            \tan \theta_{n_q - 2} | 1 \rangle
        \right)
        \otimes
        \cdots
        \otimes
        \left(
            | 0 \rangle
            +
            \tan \theta_0 | 1 \rangle
        \right)
    \nonumber \\
    &=
        C
        \sum_{q_{n_q - 2} = 0}^1
        \cdots
        \sum_{q_0 = 0}^1
            (\tan \theta_{n_q - 2})^{q_{n_q - 2}}
            \cdots
            (\tan \theta_0)^{q_0}
            | q_{n_q - 1} \rangle
            \otimes
            \cdots
            \otimes
            | q_0 \rangle
    \nonumber \\
    &=
        C
        \sum_{q_{n_q - 2} = 0}^1
        \cdots
        \sum_{q_0 = 0}^1
            \exp
            \left(
                -a
                \left(
                2^{n_q - 2}
                q_{n_q - 2}
                +
                \cdots
                +
                2^0
                q_0
                \right)
            \right)
            | q_{n_q - 2} \rangle
            \otimes
            \cdots
            \otimes
            | q_0 \rangle
    \nonumber \\
    &=
        C
        \sum_{j = 0}^{N/2 - 1}
            e^{-a j}
            | j \rangle_{n_q - 1}
    ,
\end{align}
where we used
Eq.~(\ref{gen_loc_state:def_rot_angle_for_exp_state})
for obtaining the second last equality.
It is easily understood that
any computational basis $| j \rangle_{n_q - 1}$
for $n_q - 1$ qubits
changes under flip of all the bits as follows:
\begin{align}
    | j \rangle_{n_q - 1}
    \xmapsto{\mathrm{Flip \ all}}
    \left| \frac{N}{2} - 1 - j \right\rangle_{n_q - 1}
    .
\end{align}
The state of the $n_q$-qubit system thus undergoes the $y$ rotation on the $(n_q - 1)$th qubit and the subsequent CNOT gates as
\begin{gather}
    \overbrace{
        | 0 \rangle
    }^{ | q_{n_q - 1} \rangle} 
    \otimes
    \left(
        C
        \sum_{j = 0}^{N/2 - 1}
            e^{-a j}
            | j \rangle_{n_q - 1}
    \right)
    \nonumber \\
    \xmapsto{R_y}
        \frac{C}{\sqrt{2}}
        \left(
            \cos \theta_{n_q - 1} | 0 \rangle
            +
            \sin \theta_{n_q - 1} | 1 \rangle
        \right)
        \otimes
        \sum_{j = 0}^{N/2 - 1}
            e^{-a j}
            | j \rangle_{n_q - 1}
    \nonumber \\
    \xmapsto{\mathrm{CNOT}}
        \frac{C}{\sqrt{2}}
        \left(
            \cos \theta_{n_q - 1}
            | 0 \rangle
            \otimes
            \sum_{j = 0}^{N/2 - 1}
                e^{-a j}
                | j \rangle_{n_q - 1}
            +
            \sin \theta_{n_q - 1}
            | 1 \rangle
            \otimes
            \sum_{j = 0}^{N/2 - 1}
                e^{-a j}
                | N/2 - 1 - j \rangle_{n_q - 1}
        \right)
    \nonumber \\
    =
        \underbrace{
            \frac{C}{\sqrt{2}}
            \cos \theta_{n_q - 1}
        }_{\equiv C'}
        \sum_{j = 0}^{N/2 - 1}
        \left(
            e^{-a j}
                | j \rangle_{n_q}
                +
            e^{-a (j + 1)}
                | N - 1 - j \rangle_{n_q}
        \right)
    \label{gen_loc_state:Slater_state}
\end{gather}
where we used
Eq.~(\ref{gen_loc_state:def_discrete_Slater})
for obtaining the last equality.
From the normalization condition,
${C'}^2 (1 + e^{-2 a}) (1 - e^{-N a})/(1 - e^{-2 a}) = 1$
must hold.
$C'$ is thus equal to $C_{S} (n_q, a),$ defined in 
Eq.~(\ref{gen_loc_state:Slater_state_normalization_const}).
This result and
Eqs.~(\ref{gen_loc_state:def_discrete_Slater}) and
(\ref{gen_loc_state:def_Slater_func_state})
mean that $U^{(\mathrm{S})}$ generates the SF state centered at the origin:
$
| S; a, 0 \rangle_{n_q} 
=
U^{(\mathrm{S})} | 0 \rangle_{n_q}.
$

\section{Properties of Lorentzian function states}

\subsection{Relation between the Slater and Lorentzian function states}
\label{appendix:Lorentzian_from_Slater}

The SF state, that is defined from
Eqs.~(\ref{gen_loc_state:def_discrete_Slater}) and
(\ref{gen_loc_state:def_Slater_func_state}),
undergoes QFT as
\begin{align}
    | S  ; a, 0 \rangle_{n_q}
    &\xmapsto{\mathcal{F}_{\mathrm{quant}}}
        C_{S} (n_q, a)
        \sum_{j = 0}^{N/2 - 1}
        \frac{1}{\sqrt{N}}
        \sum_{k = 0}^{N - 1}
        \left(
            e^{-a j}
            \exp \frac{i 2 \pi j k}{N}
                +
            e^{-a (j + 1)}
            \exp \frac{i 2 \pi (N - 1 - j) k}{N}
        \right)
        | k \rangle_{n_q}
    \nonumber \\
    &\overset{\lambda_k \equiv a - i 2 \pi k/N}{=}
        \frac{C_{S} (n_q, a)}{\sqrt{N}}
        \sum_{k = 0}^{N - 1}
        \sum_{j = 0}^{N/2 - 1}
            \Big(
                \exp (-\lambda_k j)
                +
                \exp \left( -\lambda_k^* (j + 1) \right)
            \Big)
            | k \rangle_{n_q}
    \nonumber \\
    &=
        \frac{C_{S} (n_q, a) }{\sqrt{N}}
        (1 - e^{-2 a})
        \sum_{k = 0}^{N - 1}
        \frac{1 - (-1)^k e^{-a N/2}}
        { 1 - 2 e^{-a} \cos (2 \pi k/N) + e^{-2 a} }
        | k \rangle_{n_q}
    \nonumber \\
    &=
        | L; a, 0 \rangle_{n_q}
    ,
    \label{gen_loc_state:QFT_on_Slater_state}
\end{align}
where we used
Eqs.~(\ref{gen_loc_state:def_discrete_Lorentzian}) and
(\ref{gen_loc_state:def_Lorentzian_func_state})
for obtaining the last equality.
This result means that the QFT of the SF state coincides with the LF state having the common $a.$

\subsection{Overlap between Lorentzian function states}
\label{appendix:overlap_btwn_Lorentzian_states}

Let us calculate the overlap
$
V (a, a', k_{\mathrm{c}})
\equiv
\langle L; a, k_{\mathrm{c}} |_{n_q} 
| L; a', 0 \rangle_{n_q}
$
between two LFs.
Since
$
| L; a, k_{\mathrm{c}} \rangle_{n_q}
=
T (k_{\mathrm{c}}) | L; a, 0 \rangle_{n_q}
=
\mathcal{F}_{\mathrm{quant}}
\cdot
U_{\mathrm{shift}} (k_{\mathrm{c}})
| S; a, 0 \rangle_{n_q}
,
$
we obtain
\begin{gather}
    V (a, a', k_{\mathrm{c}})
    \nonumber \\
    =
        \langle S; a, 0 |_{n_q} 
        U_{\mathrm{shift}} (k_{\mathrm{c}})^\dagger
        | S; a', 0 \rangle_{n_q}
    \nonumber \\
    =
        \sum_{j = 0}^{N - 1}
        S_j (n_q, a)
        S_j (n_q, a')
        \exp
        \left(
            i
            \frac{2 \pi k_{\mathrm{c}} }{N}
            j
        \right)
    \nonumber \\
    =
        C_S (n_q, a)
        C_S (n_q, a')
        \left(
            \sum_{j = 0}^{N/2 - 1}
            e^{-(a + a') j}
            \exp
            \left(
                i
                \frac{2 \pi k_{\mathrm{c}} }{N}
                j
            \right)
            +
            \sum_{j' = 1}^{N/2}
            e^{-(a + a') j'}
            \exp
            \left(
                i
                \frac{2 \pi k_{\mathrm{c}} }{N}
                (N - j')
            \right)
        \right)
    \nonumber \\
    =
        C_S (n_q, a)
        C_S (n_q, a')
        \frac{
            (
                1
                -
                (-1)^{k_{\mathrm{c}} }
                e^{- (a + a') N/2}
            )            
            \sinh (a + a')
        }{\cosh (a + a') - \cos (2 \pi k_{\mathrm{c}}/N)}
    ,
\end{gather}
where we used
Eqs.~(\ref{gen_loc_state:def_U_shift}) and
(\ref{gen_loc_state:def_discrete_Slater}).

\subsection{Normalization condition for an LC}
\label{appendix:norm_cond_for_an_LC_of_LFs}

By using the overlaps between LFs,
the normalization condition for the expansion coefficients in an LC in Eq.~(\ref{gen_loc_state:LC_of_LFs}) is written down as
\begin{align}
    1
    =
        | \langle \psi_{\mathrm{lc}} | \psi_{\mathrm{lc}} \rangle |^2
    =
        \sum_{\ell = 0}^{n_{\mathrm{loc}} - 1}
        d_\ell^2
        +
        2
        \sum_{\ell > \ell'}
            d_\ell
            d_{\ell'}
            V
            (a_\ell, a_{\ell'},
            k_{\mathrm{c} \ell} - k_{\mathrm{c} \ell'})
        .
\end{align}
When an unnormalized LC of LFs is given,
the factor for normalizing it can be calculated from this equation.
We stress here that the scaling of classical-computational cost for the normalization is $\mathcal{O} (n_{\mathrm{loc}}^2),$
which does not depend on the size of data register.

\section{Finding the optimal LC based on the Metropolis method}
\label{appendix:pseudocodes_for_optimal_LC}

Here we provide the pseudocodes for finding the optimal LC of LFs for a given target function
based on the procedure outlined in
Sect.~\ref{sed:finding_optimal_LC}.

\begin{algorithm}[H]
    \caption{Optimization of an LC of LFs for a target function}
    \label{gen_loc_state:alg:optimize_lin_combo}
	\begin{algorithmic}[1]
    	\Require 
    	    \Statex{Target function $\psi_{\mathrm{ideal}},$ initial centers $\boldsymbol{k}_{\mathrm{c init}}$ and initial decay rates $\boldsymbol{a}_{\mathrm{init}}$ of LFs, inverse temperature $\beta$ and iteration number $n_{\mathrm{M}}$ for the Metropolis method, iteration number $n_{\mathrm{p}}$ for parameter optimization}
		\Ensure
            \Statex{
            Optimal centers $\boldsymbol{k}_{\mathrm{c}},$ optimal decay rates $\boldsymbol{a}$, optimal expansion coefficients $\boldsymbol{d},$ maximized objective function $F$}
		\Function{OptimizeLC}{$
            \psi_{\mathrm{ideal}},
            \boldsymbol{k}_{\mathrm{c init}},
            \beta,
            n_{\mathrm{M}}, n_{\mathrm{p}}
		    $}
                \State $
                \boldsymbol{k}_{\mathrm{c}} :=
                \boldsymbol{k}_{\mathrm{c init}}
                , \
                \boldsymbol{a} :=
                \boldsymbol{a}_{\mathrm{init}}
                $
                \For{$i_{\mathrm{M}} = 0, \dots, n_{\mathrm{M}} - 1$}
                    \State $\ell := \mathrm{random\_int} (0, \dots, n_{\mathrm{loc}} - 1)$
                    \Comment{Single LF that can be moved}
                    \State $\switch \boldsymbol{k}_{\mathrm{c M}} := \boldsymbol{k}_{\mathrm{c}}, \ k_{\mathrm{cM} \ell} += \mathrm{random\_int} (-1, 1)$
                    \Comment{Trial centers}
                    \State $\boldsymbol{a}_{\mathrm{M}}, \boldsymbol{d}_{\mathrm{M}}, F_{\mathrm{M}} :=$ \textsc{OptimizeDecayRatesAndCoeffs}$(\psi_{\mathrm{ideal}}, \boldsymbol{k}_{\mathrm{c M}}, \boldsymbol{a}, n_{\mathrm{p}})$
                    \If{$i_{\mathrm{M}} == 0$ {\bf or} $\mathrm{random\_real (0, 1)} < e^{-\beta (F - F_{\mathrm{M}})}$}
                    \Comment{Decide whether the trial centers are adopted}
                        \State $
                            \boldsymbol{k}_{\mathrm{c}} := 
                            \boldsymbol{k}_{\mathrm{c M}}
                            , \
                            \boldsymbol{a} :=
                            \boldsymbol{a}_{\mathrm{M}}
                              $ 
                      \State $F := F_{\mathrm{M}}$
                    \EndIf
                \EndFor
            \State \Return $\boldsymbol{k}_{\mathrm{c}}, \boldsymbol{a}, \boldsymbol{d}, F$
        \EndFunction
	\end{algorithmic}
\end{algorithm}

\begin{algorithm}[H]
    \caption{Optimization of decay rates of expansion coefficients for fixed centers}
    \label{gen_loc_state:alg:optimize_params_and_coeffs}
	\begin{algorithmic}[1]
    	\Require 
    	    \Statex{Target function $\psi_{\mathrm{ideal}},$ centers $\boldsymbol{k}_{\mathrm{c init}}$ and initial decay rates $\boldsymbol{a}_{\mathrm{init}}$ of LFs, iteration number $n_{\mathrm{p}}$}
		\Ensure
            \Statex{Optimal decay rates $\boldsymbol{a},$ optimal expansion coefficients $\boldsymbol{d},$ maximized objective function $F_{\max}$}
		\Function{OptimizeDecayRatesAndCoeffs}{$\psi_{\mathrm{ideal}}, \boldsymbol{k}_{\mathrm{c}}, \boldsymbol{a}_{\mathrm{init}}, n_{\mathrm{p}}$}
            \State $\boldsymbol{a} := \boldsymbol{a}_{\mathrm{init}}$
            \For{$i_{\mathrm{p}} = 0, \dots, n_{\mathrm{p}} - 1$}
                \State Calculate $G$ and $S$ from $\{ | L; a_\ell, k_{\mathrm{c} \ell} \rangle_{n_q} \}_{\ell}$ and $\psi_{\mathrm{ideal}}$
                \State Solve $G \boldsymbol{c} = \lambda S \boldsymbol{c}$
                \State $\ell_{\mathrm{opt}} := \argmax_\ell F (\boldsymbol{c}_\ell, \boldsymbol{a}, \boldsymbol{k}_{\mathrm{c}}), \ \boldsymbol{d} := \boldsymbol{c}_{\ell_{\mathrm{opt}} }$
                \Comment{Optimal eigenvector}
                \State $F_{\max} := F (\boldsymbol{d})$
                \State Update $\boldsymbol{a}$ for lowering the objective function
            \EndFor
            \State \Return $\boldsymbol{a}, \boldsymbol{d}, F_{\max}$
        \EndFunction
	\end{algorithmic}
\end{algorithm}

\end{widetext}

\bibliography{ref}

\end{document}